\documentstyle[12pt,epsf]{article}
\newcommand{\Frac}[2]{\frac{\displaystyle #1}{\displaystyle #2}}
\newcommand{\kppg}{$K_L \rightarrow \pi^+ \pi^- \gamma $ }
\newcommand{\kppge}{K_L \rightarrow \pi^+ \pi^- \gamma \; }
\newcommand{\kpiggtot}{$K \rightarrow \pi \gamma \gamma $ }
\newcommand{\kggs}{$K_L \rightarrow  \gamma \gamma^* $ }
\newcommand{\kgg}{$K_L \rightarrow  \gamma \gamma $ }
\newcommand{\opt}{${\cal O}(p^2)$ }
\newcommand{\opc}{${\cal O}(p^4)$ }
\newcommand{\ops}{${\cal O}(p^6)$ }
\newcommand{\chpt}{$\chi$PT }

\begin{document}
\pagestyle{empty}
\begin{titlepage}
\begin{center}
\vspace*{-1.3cm}
\hfill  INFNNA-IV-97/26\\
\hfill  DSFNA-IV-97/26\\
\hfill {\tt hep-ph/9711210} \\
\hfill  March, 1998
\vspace*{1.5cm} \\
{\LARGE \bf  Analysis of $K_L \rightarrow \pi^+ \pi^- \gamma$ \\ 
 in Chiral Perturbation Theory ${}^*$\\ }
\vspace*{2cm}
{\large \sc Giancarlo D'Ambrosio$^{\dagger}$} $ \; \; $ and $ \; \; $
{\large \sc  Jorge Portol\'es$^{\ddagger}$ }
\vspace*{0.4cm} \\
Istituto Nazionale di Fisica Nucleare, Sezione di Napoli \\
Dipartamento di Scienze Fisiche, Universit\`a di Napoli \\
I-80125 Napoli, Italy \\ 
\vspace*{1cm} 

\begin{abstract}
We study the long-distance dominated \kppg decay at \ops in Chiral 
Perturbation Theory. The complete calculation of the \ops loop magnetic
amplitude is carried out. At this chiral order the model dependent 
part of the vector meson exchange contribution to the magnetic amplitude
is evaluated in the two models~: FM (Factorization Model) and FMV 
(Factorization Model in the Vector Couplings). We predict, in an almost 
model independent way, a slope of \kppg in the range $c \simeq -1.6$,$-1.8$,
consistently with the experimental value. We find that the
experimental result for the width of \kppg is compatible with a bigger 
breaking of the nonet symmetry in the weak vertices than previously stated. 
Thus we conclude that our analysis does not exclude an opposite sign
to the one given by the dominance of the pion pole in the poorly 
known \kgg amplitude. A complete analysis of the ${\cal O}(p^3)$ weak
Vector--Pseudoscalar--Pseudoscalar (VPP) vertex is also performed.
\end{abstract}
\end{center}
\vspace*{0.5cm}
PACS~: 12.15.-y,12.39.Fe,12.40.Vv,13.25.Es \\
Keywords~: Radiative non--leptonic kaon decays, Non--leptonic weak
hamiltonian, Chiral Perturbation Theory, Vector meson dominance.
\vspace*{0.3cm}\\
$^{\dagger}$ E-mail~: dambrosio@axpna1.na.infn.it \\
$^{\ddagger}$ E-mail~: Jorge.Portoles@uv.es \\
\vfill
\noindent * Work supported in part by HCM, EEC--Contract No. 
CHRX--CT920026 (EURODA$\Phi$NE).
\end{titlepage}
\newpage
\pagestyle{plain}
\pagenumbering{arabic}
\textheight 24cm

\section{Introduction}
\hspace*{0.5cm}
Radiative non-leptonic kaon decays provide relevant tests for the
ability of Chiral Perturbation Theory (\chpt) \cite{WE79,GL85}
to explain weak low-energy processes. \chpt is a natural framework
that embodies together an effective theory, satisfying the basic chiral
symmetry of QCD, and a perturbative expansion in masses
and external momenta that becomes the practical tool to work with. Its
success in the study of radiative non-leptonic kaon decays has been 
remarkable (see Refs.~\cite{RW93,DE95} and references therein). 
\par
Moreover
the chiral anomaly present in the Standard Model manifests mainly in the
low--energy interactions of the pseudoscalar mesons, and consequently
\chpt is the appropriate framework to study those effects. As has
already been shown \cite{EN92,BE92}, radiative kaon decays are 
sensitive to the chiral anomaly in the non-leptonic sector. The study
of these decays driven by the chiral anomaly ($K \rightarrow \pi \pi
\gamma$, $K \rightarrow \pi \pi \gamma \gamma$, etc.) has thoroughly
been carried out previously in the Refs.~\cite{EN92,BE92,EN94,DI95}
where their relevant features have been discussed.
\par
The main uncertainty still present in the 
computation of those channels is the contribution from local terms
that, in principle, appear at any order in the chiral expansion. At \opc 
vector
meson exchange (when present) has been shown to be effective in 
predicting the relevant couplings in the strong sector \cite{EG89,DR89}.
However our poor phenomenological knowledge of the weak processes involving
meson resonances translates into our ignorance about
the couplings in weak counterterms at \opc and beyond, thus models
are required. In particular the Factorization Model (FM) \cite{PR91} has 
been widely employed in this task \cite{EK93,BP93}. 
In Ref.~\cite{DP97a} we proposed an implementation of the FM in the Vector
Couplings (FMV) as a more efficient way of including the contributions
of vector mesons into the weak couplings at \ops in the processes
\kpiggtot and $K_L \rightarrow \gamma \ell^+ \ell^-$. We reached a good 
phenomenological description of both channels and showed that with the
new vector contributions included no enhancement due to $\Delta I = 
1/2$ transitions is required, in opposition to previous statements
in the FM. 
\par
The basic statement of the FMV is to use the idea of factorization in order
to construct the weak vertices involving vector mesons instead of 
implementing factorization in the strong lagrangian generated by vector
exchange once the vector degrees of freedom have already been 
integrated out. The apparently tiny difference between both realizations
has proven to be crucial in disentangling the r\^ole of the FM in uncovering
new chiral structures that were absent in the standard approach and
consequently giving a deeper understanding of the underlying physics.
\par
In the present article we present the application of these ideas to the
process \kppg . This decay brings new features to our previous work. While
in \kpiggtot and $K_L \rightarrow \gamma \ell^+ \ell^-$ only the weak
${\cal O}(p^3)$
Vector-Pseudoscalar-Photon ($VP \gamma$) vertex is involved, now in 
\kppg we also need to consider the weak ${\cal O}(p^3)$
Vector-Pseudoscalar-Pseudoscalar
($VPP$) vertex. The interest in studying this particular process has a 
twofold purpose: a) the relevance of the process by itself given the
fact that phenomenologically constitutes an appropriate
target of \chpt, and b) the dependence of this channel in the breaking
of nonet symmetry in the weak sector. Let us comment these two aspects in
turn.
\par
As studied time ago \cite{MA78} the process \kppg is driven by long--distance
contributions. The Direct Emission amplitude of the decay is dominated by
a magnetic type amplitude which slope in the photon energy in the kaon 
rest frame (to be defined properly in Section 4) has been measured 
experimentally. Even when the error is still a rough 30~\%  the slope 
seems rather big, a feature to be explained in \chpt as a  
contribution starting at \ops. Hence this fact together with our poor 
knowledge of the constant amplitude deserves a careful treatment.
At \ops chiral loops are present too, thus we have also performed the
full loop evaluation at this chiral order. It turns out to be relevant 
for the stability of our slope prediction.
\par
In Ref.~\cite{DP97a} we pointed out that the experimental slope of 
\kggs and the theoretical vector meson dominance prediction cannot
be accommodated unless the \kgg amplitude departs drastically
(change sign) from the one predicted by the nonet symmetry. 
We would like to suggest that at the origin of this problem could be a 
large breaking of the nonet symmetry in the weak vertex. A way to 
explore this possibility is to study this hypothesis in another process 
sensitive to this breaking: \kppg receives an \ops contribution due to
anomalous reducible amplitudes \cite{EN92,EN94} where the nonet breaking 
plays a major r\^ole. Therefore we would like to investigate also the
possibility of a satisfactory
simultaneous description of the processes \kgg and \kppg.
\par
In Section 2 we will remind briefly the basic features of \chpt and
its application to weak processes. Then in Section 3 we will specify 
the general characteristics and notation for \kppg and we will offer a 
short overview of the process in \chpt . The \ops contributions to
the magnetic amplitude will be collected and explained in Section 4. The
analysis of the combined observables will be carried out in Section 5 
while we will emphasize our conclusions in Section 6. Three brief appendices
complement the main text.

\section{Non-leptonic weak interactions at low energies}
\hspace*{0.5cm}
We will review in this Section the procedures of \chpt and their 
implementation in the study of non-leptonic weak processes. 
We will collect also the tools we will need in the development of our study.

\subsection{Chiral Perturbation Theory}
\hspace*{0.5cm}
\chpt \cite{WE79,GL85} is an effective quantum field theory for the study
of low energy strong interacting processes that relies in the exact 
global chiral
symmetry of massless QCD. The basic assumption is that, at low energies
($E \leq 1 \, \mbox{GeV}$), the chiral symmetry group $G \equiv SU(3)_L 
\otimes
SU(3)_R$ is spontaneously broken to the vector subgroup $SU(3)_V$, 
generating eight Goldstone bosons to be identified with the lightest
octet of pseudoscalar mesons ($\varphi_i$). These are conveniently 
parameterized by a $SU(3)$ matrix field 
\begin{equation}
U(\varphi) \, = \, \exp \left( \, \Frac{i}{F} \, \sum_{j=1}^{8} \lambda_j 
\varphi_j \, \right) \; , 
\label{eq:uphi}
\end{equation}
that transforms linearly under the chiral group G. In Eq.~(\ref{eq:uphi})
$\lambda_i$ are the $SU(3)$ Gell--Mann matrices \footnote{ Normalized
to $Tr(\lambda_i \lambda_j) = 2 \delta_{ij}$.} and $F \sim F_{\pi} \simeq
93$ MeV is the decay constant of the pion.
\par
The extension of the global chiral symmetry to a local one and the 
inclusion of external fields are convenient tools in order to work out
systematically the Green functions of interest and the construction
of chiral operators in the presence of symmetry breaking terms. A 
covariant derivative on the $U$ field is then defined as
\begin{equation}
D_{\mu} U \, = \, \partial_{\mu} U \, - \, i r_{\mu} U \, + \, 
                  i U \ell_{\mu} \; , 
\label{eq:ducov}
\end{equation}
where $\ell_{\mu} = v_{\mu} - a_{\mu}$ and $r_{\mu} = v_{\mu} + a_{\mu}$,
are the left and right external fields, respectively, in terms of the 
external vector and axial--vector fields. If only the electromagnetic 
field is considered then $\ell_{\mu} = r_{\mu} \, = \, - e Q A_{\mu}$ where
$Q \equiv diag(2/3,-1/3,-1/3)$ is the electric charge matrix of the
$u$, $d$ and $s$ quarks \footnote{This corresponds to 
$D_{\mu} \phi^{\pm} = (\partial_{\mu} \pm i e A_{\mu}) \phi^{\pm}$.}.
The explicit breaking of the chiral symmetry due to the masses
of the octet of pseudoscalars is included through the external scalar
field $s = {\cal M} + ...$ In this way an effective lagrangian can be 
constructed
as an expansion in the external momenta (derivatives of the Goldstone
fields) and masses \cite{WE79,GL85,MG84}.
\par
The leading \opt strong lagrangian is
\begin{equation}
{\cal L}_2 \, = \, \Frac{F^2}{4} \, \langle \, u_{\mu} u^{\mu} \, +
                   \, \chi_+ \, \rangle \; , 
\label{eq:str2}
\end{equation}
where $\langle \, A \, \rangle \, \equiv \, Tr(A)$ in the flavour space and
\begin{eqnarray}
u_{\mu} \, = \, i \, u^{\dagger} \, D_{\mu} \, U \, u^{\dagger} 
\; \; \;  & , & \; \;  U \, = \, u^2  \; \; , \nonumber \\
\chi_{+} \, = \, u^{\dagger} \, \chi \, u^{\dagger} \, + \, 
u \, \chi^{\dagger} \, u \; \; \; & , & \; \; 
\chi \, = \, 2 \, B_{\circ} \, ( s \, + \, i \, p) \, = \, 
2 \, B_{\circ} \, {\cal M} \, + ... \; \; , 
\label{eq:extrasym} \\
{\cal M} \, = \, diag(m_u \, , \, m_d \, , \, m_s) \; \; \; & , & \;
\; B_{\circ} \, = \, - \, \Frac{1}{F^2} \, \langle 0 | \,
\overline{u} u \, |0 \rangle \; . \nonumber
\end{eqnarray}
The even-intrinsic parity lagrangian at \opc was developed in Ref.~\cite{GL85}
and introduces 12 new coupling constants. For our work the only piece
we will need is
\begin{equation}
{\cal L}_4 \, = \, - \, i \, L_9 \, \langle \, f_+^{\mu \nu} u_{\mu} 
                  u_{\nu} \, \rangle \, + \, ... \; , 
\label{eq:op4l9}
\end{equation}
where
\begin{equation}
f_{\pm}^{\mu \nu} \, = \, u F_L^{\mu \nu} u^{\dagger} \, \pm \, 
                          u^{\dagger} F_R^{\mu \nu} u \; ,
\label{eq:fplus}
\end{equation}
and $F_{R,L}^{\mu \nu}$ are the strength field tensors associated to the
external $r_{\mu}$ and $\ell_{\mu}$ fields, respectively. From the 
experimental value of the pion charge radius one obtains $L_9^r(m_{\rho}) 
= (6.9 \pm 0.7) \times 10^{-3}$.
\par
The \opc odd-intrinsic parity lagrangian arises as a solution to the Ward
condition imposed by the chiral anomaly \cite{WW71}. The chiral anomalous
functional $Z_{an} [ U , \ell , r]$ as given by the Wess-Zumino-Witten
action (WZW) is
\begin{eqnarray}
Z_{an}[U,\ell,r]_{WZW} \, & = & - \, \Frac{i N_c}{240 \pi^2} \, 
\int_{M^5} \, d^5x \epsilon^{ijklm}\langle \Sigma_i^L \Sigma_j^L
\Sigma_k^L \Sigma_l^L \Sigma_m^L \rangle \nonumber \\
& &  
\label{eq:zulr} \\
& & - \, \Frac{i N_c}{48 \pi^2} \, 
\int \, d^4x \varepsilon_{\mu \nu \alpha \beta} \, 
( W(U,\ell,r)^{\mu \nu \alpha \beta} \, - \, 
W(I,\ell,r)^{\mu \nu \alpha \beta} \, ) \; , \nonumber
\end{eqnarray}
\begin{eqnarray}
W(U,\ell,r)_{\mu \nu \alpha \beta} \, & = & \langle
U \ell_{\mu} \ell_{\nu} \ell_{\alpha} U^{\dagger} r_{\beta} \, + \, 
\Frac{1}{4}U \ell_{\mu} U^{\dagger} r_{\nu} U \ell_{\alpha} 
U^{\dagger} r_{\beta} \, + \, i U \partial_{\mu} \ell_{\nu} \ell_{\alpha}
U^{\dagger} r_{\beta} \nonumber \\
& & \; + \, i \partial_{\mu} r_{\nu} U \ell_{\alpha} U^{\dagger} r_{\beta} \,
- \, i \Sigma_{\mu}^L \ell_{\nu} U^{\dagger} r_{\alpha} U \ell_{\beta} \, 
+ \, \Sigma_{\mu}^L U^{\dagger} \partial_{\nu} r_{\alpha} U \ell_{\beta} 
\nonumber \\
& & \; - \, \Sigma_{\mu}^L \Sigma_{\nu}^L U^{\dagger} r_{\alpha} U 
\ell_{\beta} \, + \, \Sigma_{\mu}^L \ell_{\nu} \partial_{\alpha} \ell_{\beta}
\, + \, \Sigma_{\mu}^L \partial_{\nu} \ell_{\alpha} \ell_{\beta} 
\nonumber \\
& & \; - \, i \Sigma_{\mu}^L \ell_{\nu} \ell_{\alpha} \ell_{\beta} \, 
+ \, \Frac{1}{2} \Sigma_{\mu}^L \ell_{\nu} \Sigma_{\alpha}^L \ell_{\beta}
\, - \, i \Sigma_{\mu}^L \Sigma_{\nu}^L \Sigma_{\alpha}^L \ell_{\beta} 
\, \rangle
\nonumber \\
& & \; - \; ( \, L \, \leftrightarrow \, R \, ) \; , 
\label{eq:wulr} 
\end{eqnarray}
with $N_c = 3$, $\Sigma_{\mu}^L \, = \, U^{\dagger} \partial_{\mu} U$,
$\Sigma_{\mu}^R \, = \, U \partial_{\mu} U^{\dagger}$
and $(L \leftrightarrow R)$ stands for the interchange
$U \leftrightarrow U^{\dagger}$,  
$\ell_{\mu} \leftrightarrow r_{\mu}$, 
$\Sigma_{\mu}^L \leftrightarrow \Sigma_{\mu}^R$. We notice that the WZW 
action does not include any unknown coupling.
\par
The inclusion of other quantum fields than the pseudoscalar Goldstone
bosons in the chiral lagrangian was considered in Ref.~\cite{CW69}. We are
interested in the introduction of vector mesons coupled to the 
$U(\varphi)$ and to the external fields. Let us then introduce 
the nonet of vector fields 
\begin{equation}
V_{\mu} \, = \, \Frac{1}{\sqrt{2}} \sum_{i=1}^{8} \lambda_i V_{\mu}^i \, 
                 + \, \Frac{1}{\sqrt{3}} V_{\mu}^0 \; , 
\label{eq:vmu}
\end{equation}
that transforms homogeneously under the chiral group $G$. Here ideal mixing,
i.e. $V_{\mu}^8 = (\omega_{\mu} + \sqrt{2} \phi_{\mu} ) / \sqrt{3}$, is 
assumed.
\par
The most general strong and electromagnetic lagrangian, at leading 
${\cal O} (p^3)$ and assuming nonet symmetry, with the vector field 
linearly coupled to the Goldstone bosons, is given by the following terms
\cite{EGL89}
\begin{eqnarray}
{\cal L}_{V} \, &  =  & \, - \, \Frac{f_V}{2 \sqrt{2}} \, \langle \, 
                             V_{\mu \nu} f_+^{\mu \nu} \, \rangle \, 
                             - \, i \Frac{g_V}{2 \sqrt{2}} \, \langle \, 
                             V_{\mu \nu} [ u^{\mu} , u^{\nu} ] \, \rangle
                             \, + \,  h_V \, \varepsilon_{\mu \nu \rho \sigma}
                       \, \langle \, V^{\mu} \{ u^{\nu}, f_+^{\rho \sigma} \}
                       \, \rangle  \nonumber \\
& & \nonumber \\
 &  & \, + \, i \, \theta_V \, \varepsilon_{\mu \nu \rho \sigma}
                      \, \langle \, V^{\mu} u^{\nu} u^{\rho} u^{\sigma} \, 
                      \rangle \; ,
\label{eq:vgpp}
\end{eqnarray}
where only the relevant pieces for our work have been written. In 
Eq.~(\ref{eq:vgpp}) $V_{\mu \nu} = \nabla_{\mu} V_{\nu}
- \nabla_{\nu} V_{\mu}$ and $\nabla_{\mu}$ is the covariant derivative 
defined in Ref.~\cite{EG89} as
\newpage
\begin{eqnarray}
\nabla_{\mu} V_{\nu} & = & \partial_{\mu} V_{\nu} \, + \, 
                        \left[ \Gamma_{\mu} , V_{\nu} \right] \; ,
\nonumber \\
& & \label{eq:cova} \\
\Gamma_{\mu} & = & \Frac{1}{2} \left\{ \, u^{\dagger} ( \partial_{\mu} 
                  - i r_{\mu} ) u \, + \, u ( \partial_{\mu} - i \ell_{\mu})
                    u^{\dagger} \, \right\} \; . \nonumber 
\end{eqnarray}
The couplings in Eq.~(\ref{eq:vgpp}) can be
determined phenomenologically from the experiment \cite{PDG96}. The 
experimental width $\Gamma(\omega \rightarrow \pi^0 \gamma)$ gives
$|h_V| \simeq 0.037$ \cite{EP90}. Also from $\Gamma (\rho^0 \rightarrow 
e^+ e^-)$ one gets $|f_V| \simeq 0.20$ and from $\Gamma (\rho^0 \rightarrow 
\pi^+ \pi^-)$ $|g_V| \simeq 0.09$ \cite{EG89}. Moreover the
positive slope of the $\pi^0 \rightarrow \gamma \gamma^*$ form factor
determined experimentally imposes $h_V f_V > 0$. The resonance saturation
of the \opc couplings gives \cite{EG89} $L_9 = f_V g_V / 2$ in 
Eq.~(\ref{eq:op4l9}) and the positivity, determined phenomenologically, of 
$L_9^r (m_{\rho})$ implies $f_V g_V > 0$. In the Hidden Gauge Model
\cite{BK88} the relations $f_V = 2 g_V$ and $\theta_V = 2 h_V$ arise. 
The first one is in excellent agreement with the experimental determination
while the second is in good agreement with the ENJL model results \cite{PR94}. 
We will assume them throughout the paper.
\par
The introduction of the axial--vector nonet defined analogously to the vector
in Eq.~(\ref{eq:vmu}) is similar to the latter. The interaction terms
we will need are
\begin{equation}
{\cal L}_{A} \, = \, - \, \Frac{f_A}{2 \sqrt{2}} \, \langle \, 
A_{\mu \nu} f_{-}^{\mu \nu} \, \rangle \, + \, h_A \, 
\varepsilon_{\mu \nu \alpha \beta} \, \langle \, A^{\mu} \, \{ u^{\nu},
f_{-}^{\alpha \beta} \} \, \rangle~,
\label{eq:agpp}
\end{equation}
with $A_{\mu \nu}$ defined as in the vector case above. The couplings
$f_A$ and $h_A$ could be determined from radiative decays. Due to the 
poor phenomenology available on axial-vector decays we rely on
theoretical predictions \cite{EGL89,PR94} that give $f_A \simeq 0.09$ and 
$h_A \simeq 0.014$.
\par
The incorporation of spin--1 mesons in chiral lagrangians is not unique
and several realizations of the vector field can be employed \cite{BI96}.
In Ref.~\cite{EGL89} was proven that at \opc in \chpt and once high energy
QCD constraints are taken into account, the usual realizations
(antisymmetric tensor, vector field, Yang-Mills and Hidden formulations) are
equivalent. Although the antisymmetric tensor formulation of spin-1 mesons
was proven to have a better high-energy behaviour than the vector field
realization at \opc, this fact is not necessarily the case in general.
In fact for the odd-intrinsic parity operator relevant in the 
$V \rightarrow P \gamma$ decay, the antisymmetric tensor formulation  
would give contributions starting at \opc while QCD requires an explicit 
${\cal O} (p^3)$ term as given by the vector realization in the $h_V$ term
in Eq.~(\ref{eq:vgpp}) \cite{EP90} \footnote{For a more detailed discussion 
of the
equivalence of vector resonance models in the odd--intrinsic parity violating
sector see Ref.~\cite{PP93}.}. Moreover we have pointed
out that the conventional vector formulation seems to give a complete and 
consistent treatment of the spin--1 resonance generated weak couplings at 
\opc \cite{DP97b} and \ops \cite{DP97a}.
\par
For a further extensive and thorough exposition on \chpt see 
Ref.~\cite{DEP92}.

\subsection{Non-leptonic weak interactions in \chpt}
\hspace*{0.5cm}
At low energies ($E \ll M_W$) the $\Delta S = 1$ effective hamiltonian
is obtained from the Lagrangian of the Standard Model by using the 
asymptotic freedom property of QCD in order to integrate out the 
fields with heavy masses down to scales $\mu < m_c$. It reads \cite{GA74}
\begin{equation}
{\cal H}_{NL}^{|\Delta S| = 1} \, = \, - \, \Frac{G_F}{\sqrt{2}} \, 
                V_{ud} V_{us}^* \, \sum_{i=1}^{6} \, C_i(\mu) Q_i \, 
                + \, h.c. \; . 
\label{eq:heffd}
\end{equation}
Here $G_F$ is the Fermi constant, $V_{ij}$ are elements of the CKM matrix
and $C_i(\mu)$ are the Wilson coefficients of the four-quark operators
$Q_i$, $i=1,...6$. ${\cal H}_{NL}^{|\Delta S| = 1}$ induces $\Delta I = 
1/2$ and $ \Delta I = 3/2$ transitions. As the first components are
phenomenologically dominant in the non--leptonic kaon decays and since
data in the structure dependent amplitude of $K_L \rightarrow \pi^+ 
\pi^- \gamma$ are accurate only until $20 \%$ level, the $\Delta I = 
3/2$ transitions can be neglected for the present purposes. Therefore we
 only will consider
the octet component of ${\cal H}_{NL}^{|\Delta S| = 1}$.
\par
At \opt in $\chi$PT we can construct only one relevant octet 
effective operator using the left-handed currents associated to the
chiral transformations,
\begin{eqnarray}
{\cal L}_2^{|\Delta S|=1} & = & 4 \, \Frac{G_F}{\sqrt{2}} V_{ud} V_{us}^*
        \, g_8  \, \langle \, \lambda_6 L_1^{\mu} L_{\mu}^1 
        \, \rangle \nonumber \\ 
& = & \Frac{G_F}{\sqrt{2}} \, F^4 \, V_{ud} V_{us}^* \, g_8 \, 
       \langle \, \Delta u_{\mu} u^{\mu} \, \rangle~,
\label{eq:weaklabare}
\end{eqnarray}
where
\begin{equation}
L_{\mu}^1 \, = \, \Frac{\delta S_2^{\chi}}{\delta \ell^{\mu}} \, = \, 
             - i \Frac{F^2}{2} U^{\dagger} D_{\mu} U \, = \, 
             - \Frac{F^2}{2} \, u^{\dagger} u_{\mu} u \; , 
\label{eq:leftcu}
\end{equation}
is the left-handed current associated to the $S_2^{\chi}$ action of the
\opt strong lagrangian in Eq.~(\ref{eq:str2}),  $\Delta = u \lambda_6
u^{\dagger}$ and $g_8$ is an effective coupling. 
\par
From the experimental width of $K \rightarrow \pi \pi$ one gets
at \opt
\begin{equation}
|g_8|_{K \rightarrow \pi \pi} \simeq 5.1 \; \; \; \; \; , \; \; \; 
\; \; G_8 \, \equiv \, \Frac{G_F}{\sqrt{2}} V_{ud} V_{us}^* \, 
|g_8|_{K \rightarrow \pi \pi} \, \simeq \, 9.2 \times 10^{-6} \, 
\mbox{GeV}^{-2} \; . 
\label{eq:g8fr}
\end{equation}
In our analysis of \kppg the singlet pseudoscalar $\eta_0$ contributes at
\ops. If nonet symmetry is broken weak interactions of the singlet at 
${\cal O}(p^2)$ have to be parameterized through a new coupling~:
$\rho$. Then  
\begin{equation}
{\cal L}_{\eta'}^{|\Delta S|=1} \, = \, \Frac{2}{3} \, G_8 \, F^4 \,
 (\, \rho \, - \, 1\, ) \, \langle \, \Delta u_{\mu} \, \rangle 
\, \langle \, u^{\mu} \, \rangle~, 
\label{eq:weakla}
\end{equation}
should be added to ${\cal L}_2^{|\Delta S|=1}$ in Eq.~(\ref{eq:weaklabare}).
\par
At \opc the weak chiral lagrangian has been studied in 
Refs.~\cite{EK93,KM90,EF91} giving 37 chiral operators only in the octet part. 
For the study of anomalous radiative decays only four of them are relevant,
\begin{eqnarray}
{\cal L}_4^{|\Delta S| =1} \, & = & \, 
G_8 F^2 \, \varepsilon_{\mu \nu \alpha \beta} \, \left[ \, 
            i \, N_{28}  \, \langle \Delta \, u^{\mu} \, \rangle \, 
                    \langle \, u^{\nu} u^{\alpha} u^{\beta} \, \rangle \, +
            N_{29} \, \langle \, \Delta \, [ f_+^{\alpha \beta} \, 
            - \, f_{-}^{\alpha \beta} \, , \, u^{\mu} u^{\nu} \, ] \,
            \rangle \, \, \right. \nonumber \\
& & \; \; \; \; \; \; \; \, \; \; \; \; \; \; \; \; \; \; \; \; \left. 
       + \, N_{30} \, \langle \, \Delta \, u^{\mu} \, \rangle \, 
                      \langle \, f_{+}^{\alpha \beta} u^{\nu} \, \rangle \,
       + \, N_{31} \, \langle \, \Delta \, u^{\mu} \, \rangle \, 
                      \langle \, f_{-}^{\alpha \beta} u^{\nu} \, \rangle \,
            \right] + ... \, , 
\label{eq:weak4}
\end{eqnarray}
where $N_{28}, \, ... \, , N_{31}$ are new coupling constants to be determined
phenomenologically or predicted by models. 
\par
Motivated by $1/N_c$ arguments the concept of factorization has been 
developed in the context of \chpt. In the $N_c \rightarrow \infty$ limit
penguin operators are suppressed. Then
if the effect of gluon exchange at leading order is considered and we
assume octet dominance through current--current operators 
in Eq.~(\ref{eq:heffd}) we have \cite{GA74}
\begin{equation}
{\cal H}_{NL}^{|\Delta S|=1} \, = \, - \, \Frac{G_F}{2 \sqrt{2}} \, 
                V_{ud} V_{us}^* \, C_{-}(\mu) \, Q_{-} \, + \, h.c. 
               \; , 
\label{eq:hds1p}
\end{equation}
where
\begin{equation}
Q_{-} \, = \, 4 \left( \, \overline{s}_L \gamma^{\mu} u_L \, \right) 
                \left( \, \overline{u}_L \gamma_{\mu} d_L \, \right) 
     \, - \,  4 \left( \, \overline{s}_L \gamma^{\mu} d_L \, \right) 
                \left( \, \overline{u}_L \gamma_{\mu} u_L \, \right) 
                \; ,
\label{eq:qm}
\end{equation}
with $\overline{q}_{1 L} \gamma_{\mu} q_{2 L} \equiv \frac{1}{2} 
\overline{q}^{\alpha}_1 \gamma_{\mu} (1 - \gamma_5) q_{2 \alpha}$ and
$\alpha$ a colour index. The $Q_{-}$ operator transforms under 
$SU(3)_L \otimes SU(3)_R$ as the $(8_L,1_R)$ representation.
The result for the Wilson coefficient $C_{-}(m_{\rho})$
at leading ${\cal O} (\alpha_s)$ and with $\Lambda_{\overline{MS}} \, 
= \, 325\,  \mbox{MeV}$ is
\begin{equation}
C_{-}(m_{\rho}) \, \simeq \, 2.2 \; \; \; \; \; \longrightarrow \; \; \; \; 
\; g_8 \, \simeq \, 1.1 \, , 
\label{eq:g81}
\end{equation}
to be compared with Eq.~(\ref{eq:g8fr}).
In a chiral gauge theory the quark bilinears in the
$Q_{-}$ operator in Eq.~(\ref{eq:qm}) are given by the associated 
left--handed current
\begin{equation}
\Frac{\delta S}{\delta \ell^{\mu}} \, = \, L_{\mu}^1 \, + \, 
L_{\mu}^3 \, + \, L_{\mu}^5 \, + ... \, ,
\end{equation} 
(the first term $L_{\mu}^1$ was already given in Eq.~(\ref{eq:leftcu}))
where $S[U,\ell,r,s,p]$ is the low--energy strong effective action of 
QCD in terms of the Goldstone bosons realization $U$ and the external
fields $\ell, r, s, p$. The assumption of Factorization 
amounts to write the four--quark operators in 
the factorized {\em current} $\times$ {\em current} form as
\begin{equation}
{\cal L}_{FM} \, = \, 4 \, k_F \, G_8 \, \langle \, \lambda \, 
\Frac{\delta S}{\delta \ell_{\mu}} \, \Frac{\delta S}{\delta \ell^{\mu}} \, 
\rangle \, + \, h.c.~,
\label{eq:fmgenera}
\end{equation}
where $\lambda \equiv \frac{1}{2} (\lambda_6 - i \lambda_7)$ and $k_F$ is
an overall factor not given by the model. In general $k_F \simeq {\cal O}(1)$
and naive factorization would imply $k_F \simeq 1$. However the Wilson 
coefficient of the $Q_{-}$ operator would give $k_F \simeq 0.2-0.3$ as
follows from Eq.~(\ref{eq:g81}) and thus a dynamical understanding of
factorization is achieved only for a value close to this one. Indeed we will
find for our factorization results this value. Therefore we imply that 
the $\Delta I = 1/2$ enhancement in $K \rightarrow \pi \pi$ is specific
of this channel but it is not a general feature of the FM in all non--leptonic
kaon decays.
\par
It has become customary in the literature \cite{BE92,EN94} to introduce
the $a_i$ couplings defined through
\begin{eqnarray}
N_{28} \, = \, \Frac{a_1}{8 \pi^2}~, \; \; & \; & \; \; \;
N_{29} \, = \, \Frac{a_2}{32 \pi^2}~, \nonumber \\
N_{30} \, = \, \Frac{ 3 \, a_3}{16 \pi^2}~, \; \; & \; & \; \; \; 
N_{31} \, = \, \Frac{a_4}{16 \pi^2}~. 
\label{eq:aidef}
\end{eqnarray} 
\par
In Refs.~\cite{BE92,CH90a} the FM was used to evaluate the factorizable 
contribution
of the chiral anomaly (WZW action in Eq.~(\ref{eq:zulr})) to these 
couplings, giving $a_i^{an} \, = \, \eta_{an}$, $i=1,2,3,4$,
with $\eta_{an} \sim {\cal O}(1)$ and positive the unknown FM factor. 
Non--factorizable contributions do not add any new chiral structure
to these.
\par
This same model predicts vanishing resonance contributions to these 
couplings if the antisymmetric formulation for the spin--1 fields
is used. In Ref.~\cite{DP97b} we 
have evaluated the spin--1 resonance contributions (vector 
and axial--vector) to
the $N_i$ couplings using a novel framework in which the vector 
formulation of the resonance fields is implemented. We quote here the
relevant results for our case using factorization. These are
\begin{eqnarray}
N_{28}^{V+A} \, & = & \, \sqrt{2} \, f_V \, \theta_V \, \eta_V~,
\nonumber \\
N_{29}^{V+A} \, & = & \, \Frac{1}{\sqrt{2}} \, ( \, f_V \, h_V \, \eta_V 
\, - \, f_A \, h_A \, \eta_A \, )~, \nonumber \\
N_{30}^{V+A} \, & = & \, 2 \sqrt{2} \, f_V \, h_V \, \eta_V~,
\nonumber \\
N_{31}^{V+A} \, & = & \, 2 \sqrt{2} \, f_A \, h_A \, \eta_A~,
\label{eq:nifmv}
\end{eqnarray}
where $f_V$, $h_V$, $\theta_V$, $f_A$ and $h_A$ have been defined in 
Eqs.~(\ref{eq:vgpp},\ref{eq:agpp}), and
$\eta_V$, $\eta_A$ are the undetermined factorization parameters.
We have, therefore, three unknown couplings~: $\eta_{an}$, $\eta_V$ and
$\eta_A$. We remind that naive factorization, however, implies 
$\eta_{an} = \eta_V = \eta_A \, \equiv \eta$. Once we determine the 
$N_i$ couplings from the phenomenology of \kppg (see Section 5) 
for sake of definiteness we will assume the naive FM relation. 
In this way we reduce the four $N_i$ couplings to only one free parameter.

\section{The \kppg amplitudes in \chpt}
\hspace*{0.5cm} 
The general amplitude for $K \rightarrow \pi \pi \gamma$ is given by
\begin{equation}
A \, [ K(p) \rightarrow \pi_1 (p_1) \pi_2 (p_2) \gamma (q, \epsilon) ] \, = \, 
\epsilon^{\mu *}(q) \, M_{\mu} (q,p_1,p_2) \; ,
\label{eq:defm}
\end{equation}
where $\epsilon_{\mu}(q)$ is the photon polarization and $M_{\mu}$ is 
decomposed into an electric $E$ and a magnetic $M$ amplitudes as
\begin{equation}
M_{\mu} \, = \, \Frac{E(z_i)}{m_K^3} \, [ \, p_1 \cdot q \, p_{2 \mu}
\, - \, p_2 \cdot q \, p_{1 \mu} \, ]
\, + \, \Frac{M(z_i)}{m_K^3} \, \varepsilon_{\mu \nu \rho \sigma} \, 
p_1^{\nu} p_2^{\rho} q^{\sigma}  \; , 
\label{eq:defem}
\end{equation}
with
\begin{eqnarray}
z_i \, = \, \Frac{q \cdot p_i}{m_K^2} \, \; \; ,(i=1,2) \; \; \; & , & 
\; \; \; \; \; \; \; z_3 \, = \, \Frac{p \cdot q}{m_K^2} \; \; \; \; \; \; , 
\; \; \; \; \; z_3 = z_1 + z_2 \; . 
\label{eq:defz}
\end{eqnarray}
The invariant amplitudes $E(z_i)$, $M(z_i)$ are dimensionless. Adding over
photon helicities there is no interference between both amplitudes and
the double differential rate for an unpolarized photon is given by
\begin{equation}
\Frac{\partial^2 \Gamma}{\partial z_1 \, \partial z_2} \, = \, 
\Frac{m_K}{(4 \pi)^3}  \left( \, |E(z_i)|^2 \, + \, |M(z_i)|^2 \, \right)
 \left[ \, z_1 z_2  (  1 - 2 (z_1+z_2) - r_1^2 - r_2^2  ) 
- r_1^2 z_2^2 - r_2^2 z_1^2 \, \right] \; , 
\label{eq:differ}
\end{equation}
where $r_i = m_{\pi_i}/m_K$. In Appendix A we recall, for completeness, 
some of the kinematical relations for \kppg.
\par
The total amplitude of $K \rightarrow \pi \pi \gamma$ can be decomposed
as a sum of {\em inner bremsstrahlung} (IB) and {\em direct emission} (DE)
(or structure dependent) amplitudes. The electric amplitude $E_{IB}(z_i)$ 
arises already at \opt in \chpt and it is completely predicted by the Low 
theorem \cite{LO58} which relates radiative and non--radiative amplitudes 
in the limit of the photon energy going to zero. Due to the pole in the
photon energy the IB amplitude generally dominates unless the 
non--radiative amplitude is suppressed due to some particular reason. 
This is the case of \kppg (where $K_2^0 \rightarrow \pi^+ \pi^-$ is 
CP violating \cite{EN94,DI95}) or $K^+ \rightarrow \pi^+ \pi^0 \gamma$ 
(where $K^+ 
\rightarrow \pi^+ \pi^0$ is suppressed by the $\Delta I = 1/2$ rule).
These channels are important in order to extract the DE amplitude 
that reveals the chiral structure of the process.
In this article we are going to focus our attention in the \kppg 
channel (for a thorough overview of other channels see Refs. 
\cite{EN94,DI95}).
\par
DE contributions can be decomposed in a multipole expansion 
\cite{DM92,LV88}.
In our following discussion we will consider only CP-conserving DE amplitudes.
These start at \opc in \chpt where E2 and M1 
multipoles are generated. Due to the asymmetry under the interchange of
pion momenta of the E2 amplitude there are no local contributions and 
then it is only generated by a finite chiral loop amplitude 
$E_{loop}^{(4)}$. It was computed in Refs.~\cite{EN92,DI95}.
The amplitude $E^{(4)}_{loop}$ gives an E2 multipole very suppressed
in comparison with the IB, typically \cite{EN94}, $|E^{(4)}_{loop} /
E_{IB}| \leq 10^{-2}$. This is due
to several circumstances: chiral loop suppression, absence of the photon
energy pole and asymmetry under the interchange of pion momenta. Higher
order terms, though in principle could be larger, still generate very
small interference with the IB amplitude \cite{DI95}. Thus we will 
neglect here the electric amplitudes.
\par
In fact the experimental results \cite{RB93} for the branching ratio 
of \kppg are consistent  with the IB ($Br(\kppge ; E_{\gamma}^* > 20 \, 
\mbox{MeV})_{IB}  =   (1.49 \pm 0.08) \times 10^{-5}$),
and a dipole magnetic contribution
\begin{equation}
Br(\kppge ; E_{\gamma}^* > 20 \, \mbox{MeV})_{DE} \,  =  \, (3.19 \pm 0.16) 
\times 10^{-5} \, ,
\label{eq:brklppg} 
\end{equation}
where $E_{\gamma}^*$ is the photon energy in the kaon rest frame.
\par
In opposition to the electric case the M1 multipole at leading \opc is 
generated by a constant local contribution only. This is given by 
${\cal L}_4^{|\Delta S|=1}$
in Eq.~(\ref{eq:weak4}) through the diagram in Fig.~1.a and the result
is
\begin{eqnarray}
M^{(4)} \, & = & \, - \, \Frac{G_8 e m_K^3}{2 \pi^2 F_{\pi}} \, 
[ \, 32 \pi^2 \, ( N_{29} + N_{31} )\, ] \nonumber \\
& = & \, - \, \Frac{G_8 e m_K^3}{2 \pi^2 F_{\pi}} \, ( \, a_2 \, + 
\, 2 a_4 \, ) \; . \label{eq:m4} 
\end{eqnarray}
\begin{figure}
\begin{center}
\leavevmode
\hbox{%
\epsfxsize=16cm
\epsffile{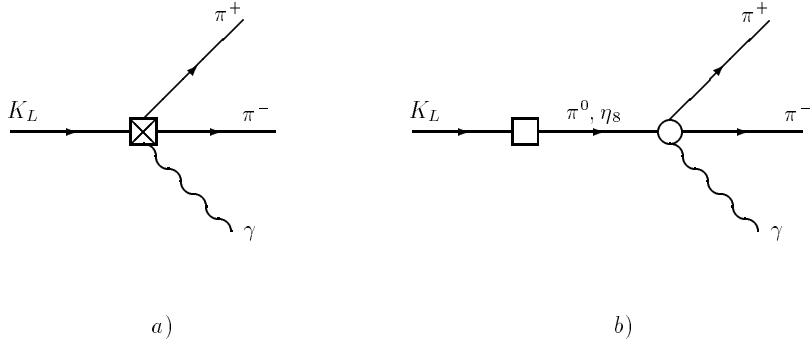}}
\end{center}
\vspace*{-15cm}
\caption{Diagrams contributing to the \opc magnetic amplitude $M^{(4)}$. The 
diagram b) is vanishing at this order. The crossed box corresponds
to a vertex generated by the ${\cal L}_4^{|\Delta S|=1}$ lagrangian
in Eq.~(\ref{eq:weak4}), the empty box is generated by 
${\cal L}_2^{|\Delta S|=1}$ in Eq.~(\ref{eq:weaklabare}) and the
circle by the WZW action $Z_{an}$ in Eq.~(\ref{eq:zulr}).}
\end{figure}
\noindent As there
is no loop contribution at this order the combination of couplings 
$N_{29} + N_{31}$ is scale independent \footnote{In fact all the 
anomalous couplings in ${\cal L}_4^{|\Delta S|=1}$ in Eq.~(\ref{eq:weak4})
are separately scale independent because there is no intrinsic parity
violating action at \opt.}. These couplings are unknown from the 
phenomenology and this fact constrains our knowledge on this
contribution. 
At \opc one could also consider the reducible anomalous
magnetic amplitude
generated by the diagram in Fig.~1.b. However at this chiral order this
amplitude vanishes because the Gell-Mann--Okubo mass relation as happens
in \kgg. Due to this cancellation and the possibility that the combination
$a_2 + 2 a_4$ is small, the \ops contributions seem theoretically important.
Also the experimental
analysis carried out in Ref. \cite{RB93} shows a clear dependence in the
photon energy that appears in the magnetic amplitude only at \ops. It 
is therefore necessary to study the \ops contributions to the magnetic
amplitude in order to be able to analyse thoroughly this decay. This we
will do in the following Sections.

\section{Magnetic amplitude of \kppg at \ops}
\hspace*{0.5cm}
Both local terms and chiral loops contribute to a magnetic amplitude for
\kppg at \ops in \chpt.
As commented before the first dependence of the magnetic amplitude in
the photon energy appears at this chiral order. 
\par 
The magnetic amplitude can be expressed as
\begin{equation}
M \, =  \, - \, \Frac{G_8 e m_K^3}{2 \pi^2 F_{\pi}} \, \widetilde{m} \, 
( \, 1 \, + \, c \, z_3 \, ) \; , 
\label{eq:slope}
\end{equation}
where $|\widetilde{m}|$ and the slope $c$ are fixed by the rate and the
spectrum, respectively \footnote{Neglecting the Electric multipoles in 
the total DE amplitude.}. From the experimental results 
\footnote{The value of the slope is given by E.J. Ramberg in private
communication reported in Ref.~\cite{EN94}. The central value can also be
followed from Ref.~\cite{RB93}. } in Ref.~\cite{RB93} we have
\begin{eqnarray}
|\widetilde{m}|_{exp} \, & = & \, 1.53 \pm 0.25 \; , \nonumber \\
& & \label{eq:expre} \\
c_{exp} \, & = & \, - \, 1.7 \pm 0.5 \; ,
\nonumber 
\end{eqnarray}
where the error in $|\widetilde{m}|_{exp}$ comes only from the error in 
the slope. $z_3$ has been defined in 
Eq.~(\ref{eq:defz}) and in the $K_L$ rest frame $z_3 = E_{\gamma}^* / m_K$.
\par
As can be noticed the dependence on the photon energy is by no means
negligible. However we have seen that it is vanishing at \opc 
in Eq.~(\ref{eq:m4}) and then it is a goal of the higher orders
in \chpt to accommodate such a big value.

\subsection{Local amplitudes}
\hspace*{0.5cm}
There are different local contributions to \kppg starting at \ops. Some of 
them are model independent while in some other cases one has to use
models. Here we do an analysis of all of them at this chiral order.
\par
First of all there is a reducible anomalous amplitude given by the 
diagram in Fig.~2. This is the same that the vanishing one in 
Fig.~1.b but now the $\pi^0$, $\eta$ and $\eta'$ are included. Then it is
\ops (but also contains higher orders) and is given by
\begin{equation}
M^{(6)}_{anom} \, = \, \Frac{G_8 e m_K^3}{2 \pi^2 F_{\pi}} \, F_1 \; , 
\label{eq:m6anom}
\end{equation}
where
\begin{eqnarray}
F_1 \, & = & \, \Frac{1}{1-r_{\pi}^2} \, + \, 
\Frac{1}{3(1-r_{\eta}^2)} \, [ (1+\xi) \cos \theta + 2 \sqrt{2} \rho
\sin \theta ] \, \left[ \left(\Frac{F_{\pi}}{F_8}\right)^3 \cos \theta - 
\sqrt{2} \left(\Frac{F_{\pi}}{F_0}\right)^3 \sin \theta \right] \nonumber \\
& & \, - \, \Frac{1}{3(1-r_{\eta'}^2)} \, [ 2 \sqrt{2} \rho \cos \theta
- (1+\xi) \sin \theta ] \, \left[\left(\Frac{F_{\pi}}{F_8}\right)^3 
\sin \theta + \sqrt{2} \left(\Frac{F_{\pi}}{F_0}\right)^3 \cos \theta \right]
\; .
\label{eq:m6f1}
\end{eqnarray}

\begin{figure}
\begin{center}
\leavevmode
\hbox{%
\epsfxsize=16cm
\epsffile{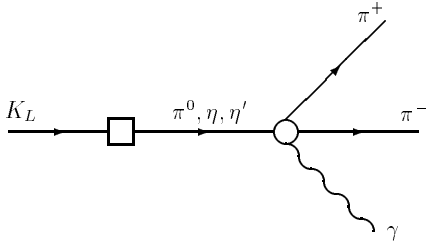}}
\end{center}
\vspace*{-16cm}
\caption{Diagram contributing to the reducible anomalous \ops magnetic 
amplitude $M^{(6)}_{anom}$. The empty box is generated by 
${\cal L}_2^{|\Delta S|=1} \, + \, 
{\cal L}_{\eta'}^{|\Delta S|=1}$ in Eqs.~(\ref{eq:weaklabare},
\ref{eq:weakla}) and the white
circle by the WZW action $Z_{an}$ in Eq.~(\ref{eq:zulr}).}
\end{figure}

\noindent In $F_1$, $r_{P} \equiv m_P / m_K$, $\xi$ parameterizes the $SU(3)$
breaking defined through
\begin{equation}
\Frac{\langle \eta_8 | {\cal L}^{|\Delta S|=1} | K_2^0 
\rangle}{\langle \pi^0 | {\cal L}^{|\Delta S|=1} | K_2^0 \rangle} \, 
= \, \Frac{1}{\sqrt{3}} \, ( 1+ \xi) \; , 
\label{eq:su3bre}
\end{equation}
$\theta$ denotes the $\eta$-$\eta'$ mixing angle, $\rho$ has been 
defined in Eq.~(\ref{eq:weakla}) and $F_8$, $F_0$ are the decay constants
of $\eta_8$ and $\eta_0$, respectively.
At \opc $F_1$ vanishes because the
Gell-Mann--Okubo mass relation. We note that $M^{(6)}_{anom}$ is constant
and has no dependence on the photon energy.
\par
Other \ops local amplitudes can be generated through resonance 
interchange. In particular we are going to focus in the vector meson
contributions which are reasonably thought to be the most relevant 
ones. There are two different kinds of these contributions 
\cite{DP97a,EP90}~:
\begin{itemize}
\item[a)] Vector exchange between strong/electromagnetic vertices with
a weak transition in an external leg. These are usually called {\em 
indirect} amplitudes and they are model independent due to our good knowledge
of the strong and electromagnetic processes involving vector mesons.
\item[b)] The {\em direct} transitions are those where the weak vertices
involving resonances are present and our poor knowledge of weak decays
of resonances makes
necessary the use of models in order to predict them.
\end{itemize}
The {\em indirect} model independent vector contribution corresponds
to the diagrams in Fig.~3 that give (assuming $m_{\pi}=0$)
\begin{eqnarray}
M^{(6)}_{ind} \, & = & \, - \, \Frac{G_8 e m_K^3}{2 \pi^2 F_{\pi}} \, 
r_V \, \left[ \, 1 \, - \, 3 \, z_3 \, \right]
\; , \nonumber \\
& & \label{eq:m6ind} \\
r_V \, & = & \, \Frac{64 \sqrt{2} \pi^2 g_V h_V m_K^2}{3 m_V^2} \, 
\simeq \, 0.41 \; , \nonumber
\end{eqnarray}
where the numerical value
of $r_V$ is for $m_V = m_{\rho}$ \footnote{If we include
the singlet $\eta_0$ and $\eta$-$\eta'$ mixing in Fig.~3 we would have 
found an extra term
\begin{equation}
\widetilde{M}_{ind}^{(6)} = \Frac{G_8 e m_K^3}{2 \pi^2 F_{\pi}} r_V F_1
\left[ \Frac{3}{2} - 3 z_3 \right]~,
\label{eq:m68}
\end{equation}
to be added to $M_{ind}^{(6)}$ in Eq.~(\ref{eq:m6ind}) ($F_1$ has been
defined in Eq.~(\ref{eq:m6f1})). However this term 
starts to contribute at ${\cal O}(p^8)$ and therefore is out of our scope
here.}. The 
strong/electromagnetic vertices present in the diagrams in Fig.~3
are given by the lagrangian density ${\cal L}_{V}$ in 
Eq.~(\ref{eq:vgpp}).
Several points are worth to comment: a) $M^{(6)}_{ind}$ shows a 
dependence in $z_3$ (i.e. in the photon energy); b) contrarily to 
what happened in the reducible anomalous contributions (Fig.~1.b and 
Fig.~2), even considering only the $\pi^0$ and the $\eta_8$ states, 
$M^{(6)}_{ind}$ does not vanish and our result 
agrees with the one quoted in Ref.~\cite{EN94}.

\begin{figure}
\begin{center}
\leavevmode
\hbox{%
\epsfxsize=16cm
\epsffile{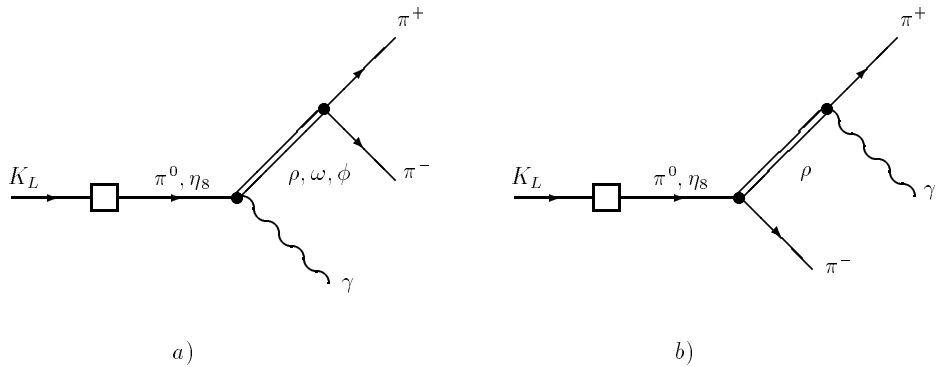}}
\end{center}
\vspace*{-15cm}
\caption{Diagrams contributing to the {\em indirect} vector meson exchange 
magnetic 
amplitude $M^{(6)}_{ind}$. The empty box is generated by 
${\cal L}_2^{|\Delta S|=1}$ in Eq.~(\ref{eq:weaklabare}) and the black circles
by the strong/electromagnetic lagrangians ${\cal L}_{V}$ in 
Eq.~(\ref{eq:vgpp}).
In b) the crossed diagram $\pi^+ \leftrightarrow \pi^-$ has to be considered
too.}
\end{figure}

Both contributions $M^{(6)}_{anom}$ and $M^{(6)}_{ind}$ are characterized
by a weak transition in the external legs given by 
${\cal L}_2^{|\Delta S|=1}$ in Eq.~(\ref{eq:weaklabare}).
We do not consider the Feynman
diagrams where this transition happens in a pion final leg because they are
suppressed by $m_{\pi}^2/m_K^2$ over those where the weak vertex
happens in the initial $K_L$ leg.
\par
The {\em direct} resonance exchange contributions are generated by 
the diagrams in Fig.~4 where a direct weak vertex involving vector
mesons is present. Due to our ignorance on these vertices from the 
phenomenological point of view it is necessary to invoke models in 
order to predict them. Moreover it has been shown \cite{DP97a,EP90}
that these contributions are not small compared with the {\em indirect}
ones and therefore have to be taken into account.
\par
Thus we construct the most general weak ${\cal O}(p^3)$ 
Vector-Pseudoscalar-Photon
($VP\gamma$) and ${\cal O}(p^3)$ Vector-Pseudoscalar-Pseudoscalar (VPP) 
vertices
appearing in the diagrams in Fig.~4 and then predict their couplings
in the FMV model.

\begin{figure}
\begin{center}
\leavevmode
\hbox{%
\epsfxsize=16cm
\epsffile{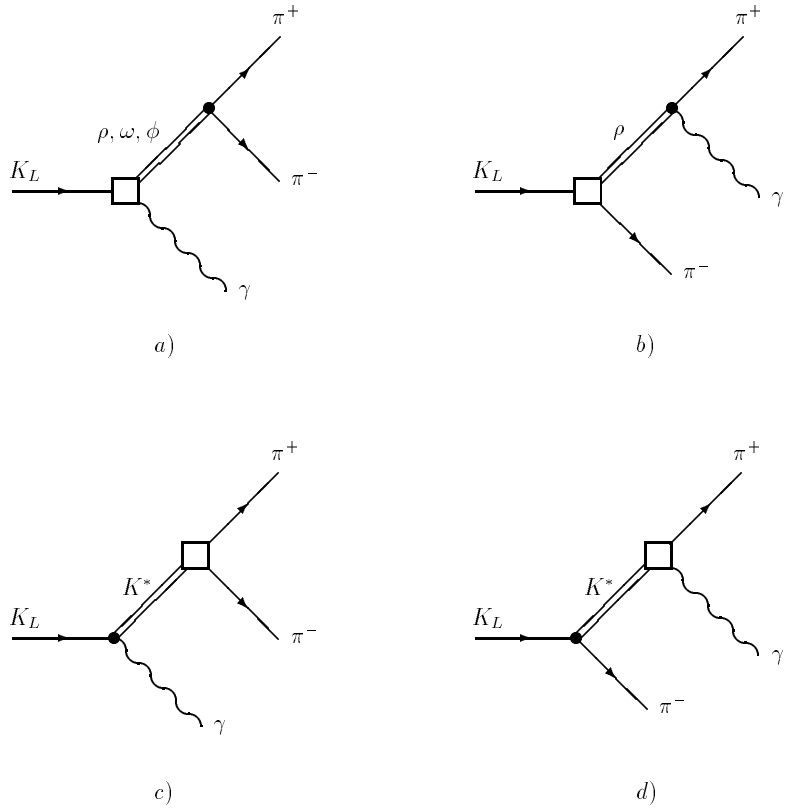}}
\end{center}
\vspace*{-6cm}
\caption{Diagrams contributing to the {\em direct} vector meson exchange 
magnetic amplitude $M^{(6)}_{dir}$. The empty box is generated by 
${\cal L}_W(VP\gamma)$ in Eq.~(\ref{eq:mostg}) or 
${\cal L}_W(VPP)$ in Eq.~(\ref{eq:mostgvpp}) and the black circles
by the strong/electromagnetic lagrangians ${\cal L}_{V}$ in 
Eq.~(\ref{eq:vgpp}).
In b) and d) the crossed $\pi^+ \leftrightarrow \pi^-$ diagrams  
are also understood.}
\end{figure}

The most general ${\cal O}(p^3)$ weak octet $VP\gamma$ vertex for the 
processes of our
interest has been already worked out in Ref.~\cite{DP97a} and it is
\begin{equation}
{\cal L}_W(VP\gamma) \; = \; G_8 \, F_{\pi}^2 \, \langle \, V^{\mu}
\, {\cal J}_{\mu}^W \, \rangle \; ,
\label{eq:mostg}
\end{equation}
with
\begin{equation}
{\cal J}_{\mu}^W \; = \; \varepsilon_{\mu \nu \alpha \beta} \; 
\sum_{j=1}^{5} \, \kappa_j \, T_j^{\nu \alpha \beta} \; ,
\label{eq:fullcur}
\end{equation}
and
\begin{eqnarray}
T_1^{\nu \alpha \beta} \; & = & \;  \{ \, u^{\nu} \, , \, \Delta \, 
                        f_{+}^{\alpha \beta} \, \} \; , 
\nonumber  \\
T_2^{\nu \alpha \beta} \; & = & \;  
          \{ \, \{ \Delta \, , \, u^{\nu} \, \} \, , \, 
                         f_{+}^{\alpha \beta} \, \} \; ,  
\nonumber \\
T_3^{\nu \alpha \beta} \; & = & \; 
      \langle \, u^{\nu} \, \Delta \, \rangle \, 
                        f_{+}^{\alpha \beta}   \; , 
\nonumber \\
T_4^{\nu \alpha \beta} \; & = & \; 
      \langle \, u^{\nu} \, f_{+}^{\alpha \beta} \, 
                      \rangle \, \Delta  \; , 
\nonumber  \\
T_5^{\nu \alpha \beta} \; & = & \; 
     \langle \, \Delta \, u^{\nu} \, f_{+}^{\alpha \beta} \, \rangle 
\; .
\label{eq:fullcurr1}
\end{eqnarray}
In the same way the most general ${\cal O}(p^3)$ weak octet VPP vertex, 
without including mass terms, is
\begin{equation}
{\cal L}_W(VPP) \, = \, G_8 \, F_{\pi}^2 \, \langle \, 
V^{\mu \nu} \, {\cal J}_{\mu \nu}^W \, \rangle \; , 
\label{eq:mostgvpp}
\end{equation}
with
\begin{equation}
{\cal J}_{\mu \nu}^W \, = \, \sum_{j=1}^{3} \, \sigma_j \, S_{\mu \nu}^j
\; , 
\label{eq:fullcurvpp}
\end{equation}
and
\begin{eqnarray}
S_{\mu \nu}^1 \, & = & \, i \, \{ \, u_{\mu} u_{\nu} \, , \, \Delta \, \} \; 
, \nonumber \\
S_{\mu \nu}^2 \, & = & \, i \, u_{\mu} \, \Delta \, u_{\nu} \; , 
\label{eq:svpp} \\
S_{\mu \nu}^3 \, & = & \, i \, \langle \, \Delta \, u_{\mu} u_{\nu} \, 
\rangle \; . \nonumber
\end{eqnarray}
In Eqs.~(\ref{eq:fullcur}) and (\ref{eq:fullcurvpp}), the $\kappa_i$ and
$\sigma_i$ are the {\em a priori} unknown coupling constants to be 
predicted in a model or given by the phenomenology when available.
\par
With the weak vertices in ${\cal L}_W(VP\gamma)$ and ${\cal L}_W(VPP)$
and the strong/electromagnetic vertices given by ${\cal L}_V$ in 
Eq.~(\ref{eq:vgpp}) we can compute now the diagrams in Fig.~4 and 
we get the direct contribution to the
magnetic amplitude as given by vector meson dominance in terms of 
the $\kappa_i$ and $\sigma_i$ couplings
\begin{eqnarray}
M^{(6)}_{dir} \, & = & \, \Frac{16}{3 \sqrt{2}} \, 
\Frac{G_8 e m_K^5}{m_V^2 F_{\pi}} \, \biggl\{  \, 
g_V \, \left[ \, 2 \kappa_2 \, + \, 3 \kappa_3 \, - \, ( 5 \kappa_2 \, + \, 
6 \kappa_3) \, z_3 \, \right] \biggr. \nonumber \\
& & \; \; \; \; \; \; \; \; \; \; \; \; \; \; \; \; \; \; \; \; \; \; \; \; 
\biggl. - \, \sqrt{2} \, h_V \, \left[ \, 2 \sigma_1 \, - \, 
( 5 \sigma_1 \, + \, \sigma_2) \, z_3 \, \right] \, \biggr\}~. 
\label{eq:m6dirmi}
\end{eqnarray}
\par
In Ref.~\cite{DP97a} we have proposed a Factorization Model in the
Vector couplings that has proven to be very efficient in the understanding
of the vector meson contributions to the \kpiggtot and \kggs processes.
The prediction for $\kappa_i$ in the FMV model has been worked out in 
Ref.~\cite{DP97a}. To evaluate the $\sigma_i$ we proceed analogously by
applying factorization to 
construct the weak VPP vertex. A detailed evaluation is shown in 
Appendix B. The predictions for the coupling constants in this model
are
\begin{eqnarray}
\kappa_1^{FMV} \, = \, 0 \; \; \; \; \; \; 
\; \; \; \; \;\; \; \; \; \;\; \; \; \; \; \; \; & ; & \nonumber \\
\kappa_2^{FMV} \, = \, ( 2 h_V - \ell_V ) \, \eta_{VP\gamma}  
& ; & \; \; \; \; \; \; \; \sigma_1^{FMV} \, = \, \; \; \; \; 
\Frac{f_V}{\sqrt{2}} \left( \eta_V + 4 L_9 
\Frac{m_V^2}{F^2} \, \eta_{VPP} \right)  \; \; , \nonumber \\
\kappa_3^{FMV} \, = \, - \Frac{8}{3} h_V \, \eta_{VP\gamma} \; \; \; 
\; \; \; \, \, 
& ; & \; \; \; \; \; \; \; \sigma_2^{FMV} \, = \, - \, \sqrt{2} 
f_V \, \eta_{V} \; \;,
\label{eq:kasig} \\
\kappa_4^{FMV} \, = \, 2 \ell_V \, \eta_{VP\gamma} \;  \; \; \; \; \;
\;  \; \; \; \;  & ; & \; \; \; \; \; \; \;  
\sigma_3^{FMV} \, = \, - \, \sqrt{2} \, f_V \, \left( \eta_V + 2 L_9 
\Frac{m_V^2}{F^2} \, \eta_{VPP} \right)  \; \; , \nonumber \\
\kappa_5^{FMV} \, = \, 2 \ell_V \, \eta_{VP\gamma} \;  \; \; \; \; \;
\; \; \;  \; \, & ; & 
\nonumber
\end{eqnarray}
where 
\begin{equation}
\ell_V \, = \, \Frac{3}{16 \sqrt{2} \pi^2} f_V \Frac{m_V^2}{F^2} \, 
\simeq \, 4 h_V \;  , 
\label{eq:lv4hv}
\end{equation}
and the last identity is exact in the Hidden gauge model and also well
supported phenomenologically. As explained in Ref.~\cite{DP97a}, $\ell_V$ 
is a contribution given by the \opc WZW anomaly Eq.(\ref{eq:zulr}).
 In Eq.~(\ref{eq:kasig}) the unknown 
factorization factors are $\eta_{VP\gamma}$ that comes
from the ${\cal O}(p^3)$ weak $VP\gamma$ vertex, $\eta_{VPP}$ from the 
${\cal O}(p^3)$ weak VPP vertex and $\eta_V$ from an ${\cal O}(p)$ weak $PV$
vertex \cite{DP97b} (see also Appendix B) and that we already have met in
the predictions for $N_i^{V+A}$ in Eq.~(\ref{eq:nifmv}).
Naive factorization, however,
would put $\eta_{VP\gamma} = \eta_{VPP} = \eta_V$. In Ref.~\cite{DP97a} 
we have
shown that the phenomenology of \kpiggtot and \kggs is well described
with $\eta_{VP\gamma} \simeq 0.21$ as predicted by the naive Wilson 
coefficient in Eq.~(\ref{eq:g81}) (i.e. no enhancement of the 
$\Delta I =1/2$ transitions) and we will use this result in our 
numerical study. For $\eta_{VPP}$ and $\eta_V$ we do not have still any 
information. However we will assume $\eta_{VPP} \simeq \eta_{VP\gamma}$
because of factorization and we will leave $\eta_V$ free. This is because
while $\eta_{VPP}$ and $\eta_{VP\gamma}$ appear both at ${\cal O}(p^3)$ 
in the weak chiral lagrangian, $\eta_V$ appears at ${\cal O}(p)$ and
therefore could differ appreciably from the naive result.
\par
With the FMV predictions for the couplings $\kappa_i$ and $\sigma_i$
we can give the direct $M_{dir}^{(6)}$ amplitude in this model. Substituting
in Eq.~(\ref{eq:m6dirmi}) we get
\begin{eqnarray}
M^{(6)}_{dir,FMV} \, & = & \, - \, \Frac{G_8 e m_K^3}{2 \pi^2 F_{\pi}} 
\, r_V \, \left[ \, (\eta_{VP\gamma} + \eta_{V}) \, ( 1 - \Frac{3}{2} z_3 ) 
\, \right. 
\nonumber \\
& & \; \; \; \; \; \; \; \; \; \; \; \; \; \; \; \; \; \; \; \; \; \; \; \; 
+ \, \eta_{VP\gamma} \, \, \Frac{\ell_V}{4 h_V} \, ( 2 - 5 z_3) \, 
\label{eq:m6dir} \\
& & \; \; \; \; \; \; \; \; \; \; \; \; \; \; \; \; \; \; \; \; \; \; \; \; 
\left.
+ \, 2 \eta_{VPP} \; L_9 \Frac{m_V^2}{F^2} \, (2 - 5 z_3) \, \right] \; , 
\nonumber
\end{eqnarray}
where $r_V$ has been defined in Eq.~(\ref{eq:m6ind}), $\ell_V$ in 
Eq.~(\ref{eq:lv4hv}) and we have put $m_{\pi} = 0$. Note that 
$M^{(6)}_{dir}$ gets a dependence on the photon energy. 
\par
Another procedure to implement the hypothesis of factorization 
has also been used in the literature \cite{EN94,EP90}. In this case one
determines first the strong action generated by vector meson dominance
and then applies the FM as given by Eq.~(\ref{eq:fmgenera}). This is 
the so called procedure ${\cal A}$ \cite{EK93,DP97b} in opposition to the
FMV model where one first determines in Factorization the weak couplings
of vectors and then integrate them out (procedure ${\cal B}$). 
\par
According to the method ${\cal A}$ the vector generated \ops strong 
action necessary to compute the direct magnetic amplitude for 
\kppg has two contributions~: i) the one given by integrating the 
vector mesons between the terms in $g_V$ and $h_V$ in ${\cal L}_V$
(Eq.~(\ref{eq:vgpp})), and ii) a term obtained by integrating the 
vectors between the terms in $f_V$ and $h_V$ in Eq.~(\ref{eq:vgpp}). 
Once factorization is applied these generate the following magnetic
amplitude~:
\begin{equation}
M^{(6)}_{dir,FM} \, = \, - \, \Frac{G_8 e m_K^3}{2 \pi^2 F_{\pi}} \, 
r_V \, k_F \, \left[ \, (2 - 3 z_3) \, + \, 
 \Frac{f_V}{2 g_V} (2 - 5 z_3) \, \right]~,
\end{equation}
where $r_V$ has been defined in Eq.~(\ref{eq:m6ind}) and $k_F$ is the
factorization parameter. In $M^{(6)}_{dir,FM}$ the term proportional
to $f_V$ is the one given by the second contribution (ii) above.
In Ref.~\cite{EN94} only the term generated
by the first structure (i) has been considered.
\par
The theoretical comparison between the two models motivates the 
following considerations.
\begin{itemize}
\item[1)] As has been shown in Ref.~\cite{DP97a} the FM and FMV models
do not give the same contributions. In fact the FMV model recovers the 
results of the FM but it is able to generate a more complete set of effective
operators. The fact
that one recovers more chiral operators in the FMV model was shown to be
important in Ref.~\cite{DP97a} for the understanding of the vector meson
contributions to \kpiggtot and \kggs. In \kppg the weak VPP vertex 
generates an extra contribution (last line in Eq.~(\ref{eq:m6dir}))
that the FM does not get \cite{EN94} \footnote{In comparing our result
$M^{(6)}_{dir,FMV}$ with the one given by the FM in Ref.~\cite{EN94} 
care has to be taken because
the anomalous contribution given by the term proportional to $L_9$ in the
FM (Eq. (5.24) of Ref.~\cite{EN94}) corresponds to our term proportional to 
$\ell_V$ in Eq.~(\ref{eq:m6dir}) and coincides with it once the relation
$L_9 = f_V g_V /2 = f_V^2/4 $ is used.}.
\item[2)] The $\sigma_i^{FMV}$ couplings in the weak VPP vertex have two 
different kinds of contributions. From Eq.~(\ref{eq:kasig}) we see that 
there is a term proportional to $\eta_{VPP}$ and a piece proportional
to $\eta_V$. The first one, analogous to the $\eta_{VP\gamma}$ 
contribution in $\kappa_i^{FMV}$, comes from the application of 
factorization through the left--handed currents in the direct $VPP$ 
vertex. As explained in Appendix B the origin of the term proportional
to $\eta_V$ is a weak shift in the kinetic term of the vector fields 
in the conventional vector formulation. This framework, that we have
proposed in Ref.~\cite{DP97b}, has been able to generate new features
in the \opc $N_i$ couplings of ${\cal L}_4^{|\Delta S| = 1}$
like, for example, non--vanishing spin--1 resonance exchange contributions
to the $N_{28},...,N_{31}$ couplings in Eqs.~(\ref{eq:weak4},\ref{eq:nifmv})
that are predicted to be zero in the FM with the antisymmetric formulation
\cite{EK93}. It is 
reassuring to notice that the term proportional to $\eta_V$ allows us 
to recover the result 
of the FM \cite{EN94} for \kppg (that is ${\cal O}(p^6)$) in agreement 
with our 
claim that the procedure we are applying uncovers new structures but 
does not miss any contained in the previous FM. This 
shows that our approach in Ref.~\cite{DP97b} is also consistent at 
\ops and hence with the chiral expansion.
\end{itemize}
We think that the FMV model seems to give a more complete
description and therefore we will analyse the \kppg process in this model.
\par 
Using the relations $\ell_V = 4 h_V$ (Eq.~(\ref{eq:lv4hv})) and 
$L_9 = f_V^2/4$ we can give a simplified expression for $M^{(6)}_{dir,FMV}$,
\begin{equation}
M^{(6)}_{dir,FMV} \, = \, - \, \Frac{G_8 e m_K^3}{2 \pi^2 F_{\pi}} \, 
r_V \, \, \eta_{VP\gamma} \, \left[ \, 3 + 2 \, \omega + \omega' \, - 
\, \Frac{1}{2} \left( 13 + 10 \, \omega + 3 \, \omega' \right) \, z_3 \, 
\right] \; , 
\label{eq:m6dsimp}
\end{equation}
where $\omega = \eta_{VPP} / \eta_{VP\gamma}$,  $\omega' = \eta_V /
\eta_{VP\gamma}$ and naive factorization
would suggest $\omega = \omega' = 1$. However, as commented before, we would 
like to emphasize that while
the value of $\eta_{VP\gamma} \simeq 0.21$ has been fixed in 
Ref.~\cite{DP97a} from the phenomenology of \kpiggtot and \kggs, no similar
constraint affects $\eta_{VPP}$ or $\eta_V$ and it could happen that, in 
contradistinction with $\eta_{VP\gamma}$, these weak vertices are affected
by the $\Delta I = 1/2$ enhancement and therefore $\omega$ and $\omega'$
could differ appreciably from the unity. It is worth to notice that,
in principle, both contributions to $M^{(6)}_{dir,FMV}$ coming from the weak
$VP\gamma$ (the pure numerical term in the brackets in 
Eq.~(\ref{eq:m6dsimp})) and 
$VPP$ vertices (terms in $\omega$ and $\omega'$) are quantitatively 
comparable and of the same order. Thus we find that the
phenomenological assumption by Heiliger and Sehgal \cite{HS93} that the
vector exchange generated DE amplitude of \kppg is dominated by the weak
$VP\gamma$ vertex as in \kpiggtot and \kggs, is not supported by our
analysis. 
\par
Moreover, comparing $M^{(6)}_{dir,FMV}$ and $M^{(6)}_{ind}$ in 
Eq.~(\ref{eq:m6ind}) we also see that both {\em direct} and {\em indirect}
contributions are comparable as it happens in \kpiggtot and \kggs.

\subsection{Loop amplitudes}
\hspace*{0.5cm}
The leading loop contributions to the magnetic amplitude of \kppg
are \ops in \chpt. There are two different kinds of loop
generated amplitudes.
\par
The first type corresponds to the Feynman diagrams in Fig.~5 which 
contribution is given by the lagrangian densities ${\cal L}_2^{|\Delta S|
=1}$ in Eq.~(\ref{eq:weaklabare}), the 
strong/electromagnetic ${\cal L}_2$ in Eq.~(\ref{eq:str2}) and the
WZW anomalous action $Z_{an}[U,\ell,r]_{WZW}$ in Eq.~(\ref{eq:zulr}). Thus 
the main feature is that this contribution is completely
specified on chiral symmetry and anomaly grounds and, therefore, it
is model independent.

\begin{figure}
\begin{center}
\leavevmode
\hbox{%
\epsfxsize=16cm
\epsffile{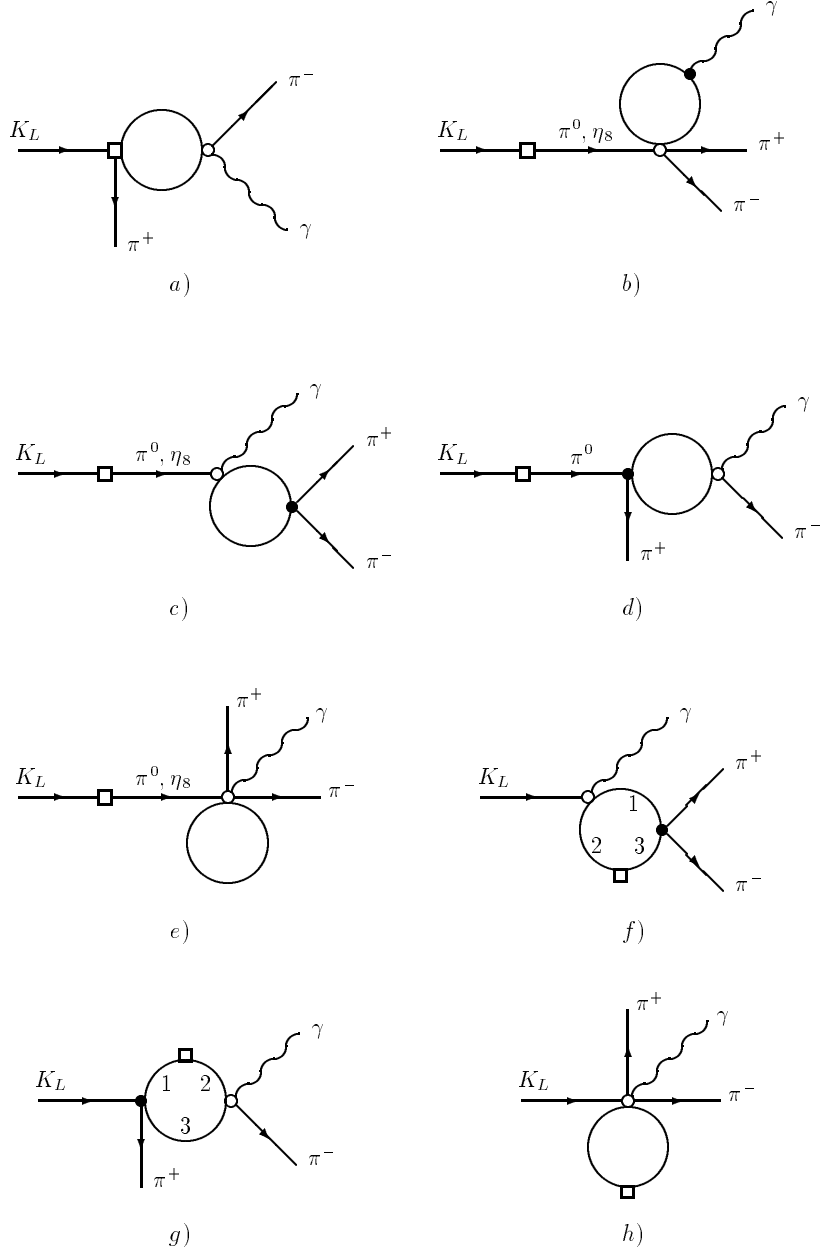}}
\end{center}
\vspace*{-4cm}
\caption{Diagrams contributing to the {\em loop} 
magnetic amplitude $M^{(6)}_{WZW}$. The empty box is generated by 
${\cal L}_2^{|\Delta S|=1}$ in Eq.~(\ref{eq:weaklabare}),
the white circles by the WZW anomalous action $Z_{an}$ in 
Eq.~(\ref{eq:zulr}) and the black circles
by the strong/electromagnetic lagrangian ${\cal L}_2$ in 
Eq.~(\ref{eq:str2}).
In a), d) and g) the crossed $\pi^+ \leftrightarrow \pi^-$ diagrams  
are also taken into account. The particle content in the loops is 
explained in the text.}
\end{figure}

In Fig.~5.a the fields running in the loop are the pairs $(\eta_8,\pi^{\pm})$
and $(\pi^0, \pi^{\pm})$. In the pole--like diagrams in Figs.~5.b--5.e we 
keep those where the external kaon leg undergoes the weak transition to 
$\pi^0, \, \eta_8$ and disregard those (suppressed by $m_{\pi}^2 / m_K^2$)
where the transition happens in a final leg. In the diagram in Fig.~5.b
only $K^+$ is running in the loop and its final contribution vanishes because
a cancellation between the $\pi^0$ and the $\eta_8$; in Fig.~5.c we have 
$\pi^+, K^+, K^0$; 
in Fig.~5.d the couple $(\pi^{\pm}, \pi^0)$ and in Fig.~5.e we have
$\pi^+, \pi^0, K^+$ and $K^0$. We note also that there is no pole
contribution with $\eta_8$ in the diagram in Fig.~5.d. In Fig.~5.f--5.h 
the weak transition happens inside the loop. In the diagram in Fig.~5.f
we have $(1,2,3) \equiv (K_S, \pi^0, K_L), (K_S, \eta_8, K_L)$ ; in 
Fig.~5.g we have (clockwise running in the loop) $(1,2,3) \equiv 
(K^-, \pi^-, \eta_8), (K^-, \pi^-, \pi^0),
(K_L, \pi^0, \pi^+), (K_L, \eta_8, \pi^+)$ and, finally, in Fig.~5.h the
$(K^+,\pi^+)$ and $(K^-,\pi^-)$ pairs appear.
\par
The result is divergent and in the $\overline{MS}$ subtraction scheme is 
\footnote{We recall that we have neglected diagrams with a weak transition 
in a external pion leg (suppressed by $r_{\pi}^2$) and these cancel the
$(1-r_{\pi}^2)$ in the denominators of $M^{(6)}_{WZW}$.}
\begin{eqnarray}
M^{(6)}_{WZW}  & = &  \Frac{G_8 e m_K^3}{2 \pi^2 F_{\pi}}  
\left( \Frac{m_K}{4 \pi F_{\pi}} \right)^2  \left[ \, 
 \Frac{5}{9} r_{\pi}^2 \, - \, \Frac{5}{9} \, + \, 
\Frac{r_{\pi}^2(7 - 9 r_{\pi}^2)}{3(1-r_{\pi}^2)}
 \, \ln \left( \Frac{m_{\pi}^2}{\mu^2} \right) \, + \,
  \Frac{7 r_{\pi}^2 - 5}{3(1-r_{\pi}^2)}
 \, \ln \left( \Frac{m_K^2}{\mu^2} \right) \right. 
\nonumber \\
& & \; \; \; \; \; \; \; \; \; \; \; \; \; \; \; \; \; \; \; \; \; \; 
\; \; \; \; \; \; \; + \; \left[ \, \Frac{5}{3} \, - \, \Frac{2}{3} \, 
\ln \left( \Frac{m_K^2}{\mu^2} \right) \, - \, \Frac{1}{3} \, 
\ln \left( \Frac{m_{\pi}^2}{\mu^2} \right) \, \right]  \, z_3 
\label{eq:m6wzw} \\
& & \; \; \; \; \; \; \; \; \; \; \; \; \; \; \; \; \; \; \; \; \; \; 
\; \; \; \; \; \; \; + \; \Frac{1}{1-r_{\pi}^2} \, \left(
 K[(p-p_+)^2, m_{\pi}^2, m_{\pi}^2] \, + \, 
K[(p-p_-)^2, m_{\pi}^2, m_{\pi}^2] \, \right) \nonumber  \\
& & \; \; \; \; \; \; \; \; \; \; \; \; \; \; \; \; \; \; \; \; \; \; 
\; \; \; \; \; \; \; - \, \Frac{1}{2(1-r_{\pi}^2)}  
\left( \, K[(p-p_+)^2, m_{K}^2, m_{\eta}^2] 
\, + \, 
 K[(p-p_-)^2, m_{K}^2, m_{\eta}^2] \, \right.  \nonumber \\
 & & \; \; \; \; \; \; \; \; \; \; \; \; \; \; \; \; \; \; \; \; \; \; 
\; \; \; \; \; \; \; \left. \; \; \; \; \; \; \; \; \; \; \; \; \; \;
\; \; \; \; \;  + \, 
  K[(p-p_+)^2, m_{K}^2, m_{\pi}^2] 
\, + \, 
 K[(p-p_-)^2, m_{K}^2, m_{\pi}^2] \, \right) \nonumber  \\ 
& & \; \; \; \; \; \; \; \; \; \; \; \; \; \; \; \; \; \; \; \; \; \; 
\; \; \; \; \; \; \; + \, \Frac{4-r_{\pi}^2}{1-r_{\pi}^2} \, 
K[(p_+ + p_-)^2,m_{\eta}^2,m_K^2] \, + 
\Frac{r_{\pi}^2}{1-r_{\pi}^2} \, K[(p_+ + p_-)^2, m_K^2,
m_{\pi}^2] \, \nonumber \\
& & \; \; \; \; \; \; \; \; \; \; \; \; \; \; \; \; \; \; \; \; \; \; 
\; \; \; \; \; \; \;  
- \, \Frac{4}{1-r_{\pi}^2} \, K[(p_+ + p_-)^2, m_K^2, m_K^2] \, 
 - \; \, 
\Frac{4}{3} \, F[(p_{+}+p_{-})^2,m_K^2] \;  \nonumber  
\\
& & \; \; \; \; \; \; \; \; \; \; \; \; \; \; \; \; \; \; \; \; \; \; 
\; \; \; \; \; \; \; \left.
+ \; \, \Frac{2}{3} r_{\pi}^2 \, \left( \, F[(p-p_{+})^2,m_{\pi}^2] \, 
+ \, F[(p-p_{-})^2,m_{\pi}^2] \, \right) \, \right] ,  \nonumber 
\end{eqnarray}
where the functions $F[q^2,m^2]$ and $K[q^2,m_1^2,m_2^2]$ have been 
defined in the Appendix C.
\par
We have compared the loop evaluation of $\pi^0 \rightarrow \pi^+ \pi^-
\gamma$ and $\eta_8 \rightarrow \pi^+ \pi^- \gamma$ (necessary for the
pole--like diagrams in Fig.~5) with the results shown in the Ref.~\cite{BB90} 
and found complete agreement. From our Eq.~(\ref{eq:m6wzw}) it is easier
to obtain the appropriate momentum dependence that is not properly 
specified in Ref.~\cite{BB90}.
Moreover the divergent part
of these diagrams comes from the strong/electromagnetic loop because
the weak transition happens in the external leg. We have checked, then, 
that the strong intrinsic parity violating \ops lagrangian \cite{AA91}
can absorb these divergences.
\par
The second type of loop contributions corresponds to the Feynman
diagrams in Fig.~6. Their basic characteristic is that the weak vertices
are generated by ${\cal L}_4^{|\Delta S|=1}$ in Eq.~(\ref{eq:weak4}) and
therefore the generated amplitudes are proportional to the, a priori
unknown, $N_i$ couplings. The strong vertices in Fig.~6 are generated by 
${\cal L}_2$ in Eq.~(\ref{eq:str2}). In Fig.~6.a the fields running in the
loop are the pairs $(\pi^+, \pi^-)$ and $(K^+, K^-)$; in Fig.~6.b we have
$\pi^+, \pi^0, \eta_8, K^0$ and $K^+$; finally in the graph in Fig.~6.c
there are the pairs $(K^0, \pi^{\pm})$, $(K^{\pm}, \pi^0)$ and 
$(K^{\pm}, \eta_8)$. We also have taken into account the field and $F_{\pi}$
renormalization due to the leading $M^{(4)}$ amplitude in 
Eq.~(\ref{eq:m4}).

\begin{figure}
\begin{center}
\leavevmode
\hbox{%
\epsfxsize=16cm
\epsffile{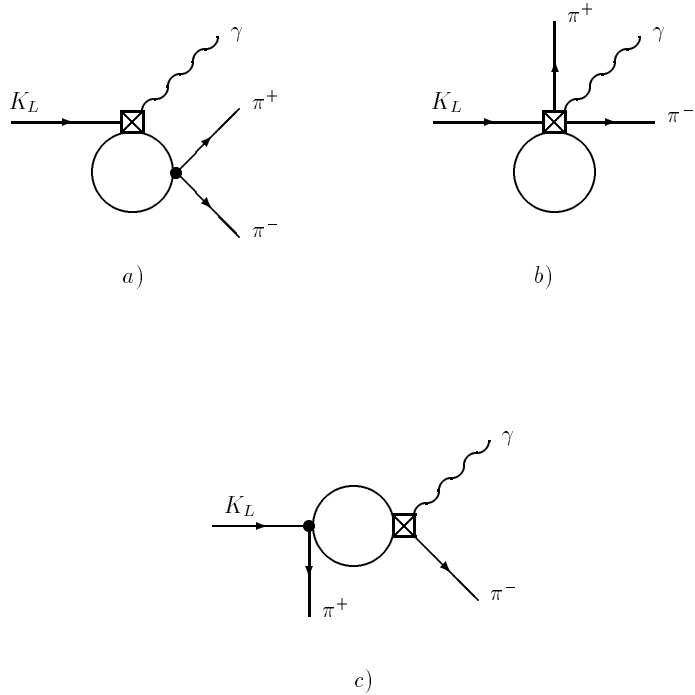}}
\end{center}
\vspace*{-7.5cm}
\caption{Diagrams contributing to the {\em loop} 
magnetic amplitude $M^{(6)}_{N_i}$. The crossed box is generated by 
${\cal L}_4^{|\Delta S|=1}$ in Eq.~(\ref{eq:weak4}),
and the black circles
by the strong lagrangian ${\cal L}_2$ in 
Eq.~(\ref{eq:str2}).
In c) the crossed $\pi^+ \leftrightarrow \pi^-$ diagram 
is also considered. The particle content in the loops is explained in 
the text.}
\end{figure}

As in the case above the result is divergent and using the $\overline{MS}$
scheme it reads

\begin{eqnarray}
M^{(6)}_{N_i} \, & = &  \, \Frac{G_8 e m_K^3}{2 \pi^2 F_{\pi}} \, 
 \left( \Frac{m_K}{4 \pi F_{\pi}}
\right)^2 \, (8 \pi^2) \; \cdot \nonumber \\
& & \;  \left\{ 3 \, N_{28} \, \ln \left( \Frac{m_K^2}{\mu^2} \right) 
\right. \;
\label{eq:m6ni} \\
& & \; \; \left. + \, (3 N_{29} - N_{30}) \, \left[ \, 2 \, \ln 
\left( \Frac{m_K^2}{\mu^2}
\right) \, + \, K[(p-p_+)^2,m_K^2,m_{\pi}^2] \right. \right. \nonumber \\
& & \; \; \; \; \; \; \; \; \; \; \; \; \; \; \; \; \; \; \; \; \; \; \; 
\; \; \; \; + \, K[(p-p_-)^2,m_K^2,m_{\pi}^2] \,  
+ \, K[(p-p_+)^2,m_K^2,m_{\eta}^2] \, \nonumber \\
& & \; \; \; \; \; \; \; \; \; \; \; \; \; \; \; \; \; \; \; \; \; \; \; \; \; 
\; \;  \biggl. + \, K[(p-p_-)^2,m_K^2,m_{\eta}^2] \,
\biggr] \, \nonumber \\ 
& & \; \; + \, 2 \, ( N_{29} + N_{31} ) \, \left[ \,
  - \, \Frac{5}{6} \, + \, \left( \Frac{15}{4} 
r_{\pi}^2 \, + \, \Frac{1}{3} \, \right) \, \ln \left(\Frac{m_{\pi}^2}{\mu^2} 
\right) \, + \, \Frac{8}{3} \, \ln \left( \Frac{m_K^2}{\mu^2} \right) \, 
\right. \nonumber \\
& & \; \; \; \; \; \; \; \; \; \; \; \; \; \; \; \; \; \; \; \; \; \; \; 
\; \; \; \; 
+ \, \Frac{3}{4} \, r_{\eta}^2 \, \ln \left( \Frac{m_{\eta}^2}{\mu^2}
\right) \,  + \, \left[  \, \Frac{5}{3} \, - \, 
\Frac{2}{3} \, \ln \left( \Frac{m_{\pi}^2}{\mu^2} \right) \, - \, 
\Frac{1}{3} \, \ln \left( \Frac{m_K^2}{\mu^2} \right) \, \right] \, z_3 
\nonumber  \\
& & \; \; \; \; \; \; \; \; \; \; \; \; \; \; \; \; \; \; \; \; \; \; \;
\; \; \; \; 
 - \, \Frac{4}{3} r_{\pi}^2 \, F[(p_{+}+p_{-})^2,m_{\pi}^2]
\, - \, \Frac{2}{3} \, F[(p_{+}+p_{-})^2,m_K^2]  \; \nonumber \\ 
& & \; \; \; \; \; \; \; \; \; \; \; \; \; \; \; \; \; \; \; \; \; \; \; 
\; \; \; \; \biggl. \biggl.
 + \, K[(p-p_+)^2, m_K^2,m_{\pi}^2] \, + K[(p-p_-)^2, m_K^2,m_{\pi}^2] 
\biggr] \biggr\} \;  , \nonumber
\end{eqnarray}
and the functions $F[q^2,m^2]$ and $K[q^2,m_1^2,m_2^2]$ have been defined in 
the Appendix C.
\par
We notice that $M^{(6)}_{N_i}$ depends on the four $N_{28}, ...., N_{31}$
unknown couplings in the \opc weak lagrangian, but only three combinations of
them, $N_{28}$, $3N_{29} - N_{30}$ and $N_{29} + N_{31}$, appear.
\par
In our numerical study of the branching ratio and spectrum in the next
section we have worked with the full expressions of $M^{(6)}_{WZW}$ and
$M^{(6)}_{N_i}$ with $\mu = m_{\rho}$. However in order to make more 
transparent our study
of the slope in \kppg (Eq.~(\ref{eq:slope})) we have linearized the
real part of the loop contributions by Taylor expanding the amplitude
assuming $4 r_{\pi}^2 \leq z_3 \ll 1$. We get for the full \ops loop
contribution
\begin{equation}
M^{(6)}_{Loop} \, = \, \Frac{G_8 e m_K^3}{2 \pi^2 F_{\pi}} \, 
\left( \Frac{m_K}{4 \pi F_{\pi}} \right)^2 \, \left[ \, a_L \, + \, 
b_L \, z_3 \, \right] \; , 
\label{eq:m6oop}
\end{equation}
where
\begin{eqnarray}
a_L \, & = & \, 0.54 \, - \, 2.63 \, a_1 \, - \, 2.07 \, (a_2 + 2 a_4) \, 
- \, 0.13 \, (a_2 - 2 a_3)~, \nonumber \\
& & \label{eq:albl} \\
b_L \, & = & \, 0.71 \, + \, 0.67 \, (a_2 + 2 a_4) \, - \, 0.07 \, 
(a_2 - 2 a_3)~. \nonumber 
\end{eqnarray}
The linearizations give values for $Re(M^{(6)}_{Loop})$ in very good 
agreement with the exact results (in a very few percent inside the domain
of $z_3$). From Eq.~(\ref{eq:albl}) we notice that if $a_i \sim {\cal O}(1)$
(as expected) the main contributions come from the terms in 
$a_1 \propto N_{28}$
and $(a_2 + 2 a_4) \, \propto \, (N_{29} + N_{31})$. 

\section{Analysis of \kppg}
\hspace*{0.5cm}
Collecting all our previous results for the magnetic amplitude of 
\kppg (with the linearized version of the loop contributions) we get
for the factor $\widetilde{m}$ and the slope $c$ defined in 
Eq.~(\ref{eq:slope}) the results
\begin{eqnarray}
\widetilde{m} & = &  a_2 \, + \, 2 a_4 \, - \, F_1 \, + \, r_V  \left[ \, 1 +  
 \eta_{VP\gamma} \, ( 3 + 2 \, \omega + \omega' ) \, \right] \, 
 - \, \left( \Frac{m_K}{4 \pi F_{\pi}} \right)^2  a_L~,
 \label{eq:mtilde} \\
& & \nonumber \\
c  & = &  - \, \Frac{1}{\widetilde{m}} \, \left[ \, r_V 
\left( \, 3 \, + \Frac{\eta_{VP\gamma}}{2} \, (13 + 10 \, \omega + 3 \, 
\omega' ) \, \right) \,  + \, \left( 
\Frac{m_K}{4 \pi F_{\pi}} \right)^2 \, b_L \, \right] \; .
\label{eq:cexpress} 
\end{eqnarray}
We recall here what is known and what is not in the expressions above:
\begin{itemize}
\item[a)] The \opc $a_2 + 2 a_4$ term is unknown. In Subsection 2.2 we 
have given the predictions for $a_i$ in the FM (anomaly) and the 
vector formulation framework using factorization (spin--1
resonances) in terms of one free parameter $\eta$. We will study our
observables in a reasonable range of this term~: $0 < a_2 + 2 a_4  \leq 3$
(we remind that factorization predicts  $a_i \sim {\cal O}(1)$ and 
positive).
\item[b)] $F_1$ has been defined in Eq.~(\ref{eq:m6f1}) and the only
parameter we will allow to be free is the one measuring the breaking of
the nonet symmetry in the weak vertex $\rho$ defined in 
Eq.~(\ref{eq:weakla}). We will consider $0 \leq \rho \leq 1$ as a
conservative working region. 
\item[c)] $r_V$ is given in Eq.~(\ref{eq:m6ind});
$\eta_{VP\gamma}$ has been fixed in Ref.~\cite{DP97a} to 
$\eta_{VP\gamma} \simeq 0.21$ as commented before.
\item[d)] The ratio $\omega = \eta_{VPP} / \eta_{VP\gamma}$, defined in 
Eq.~(\ref{eq:m6dsimp}), is also free and we will allow a domain
$0.5 \leq \omega \leq 1.5$ coherently with the naive estimate of 
factorization that sets $\omega \simeq 1$. However we will keep free
$\omega' = \eta_V / \eta_{VP\gamma}$ and will allow a reasonable
range of values for $\eta_V$, i.e. $0 < \eta_V \leq 1$.
\item[e)] The $a_L$ and $b_L$ parameters have been defined in 
Eq.~(\ref{eq:albl}) and therefore they depend on the unknown $a_i$
couplings. Our procedure will be the following. For every value of the
\opc $a_2 + 2 a_4$ free parameter we use the results in 
Eqs.~(\ref{eq:aidef},\ref{eq:nifmv}) with $\eta_V = 
\eta_{an} \equiv \eta$ to fix a value for $\eta$ as given by
\begin{equation}
\eta \, = \, \Frac{a_2 + 2 a_4}{3 + \Frac{32 \pi^2}{\sqrt{2}} f_V h_V}~.
\label{eq:eta2}
\end{equation}
With this value then we fix $a_1$ and $a_2 - 2 a_3$ that appear in the loop
contributions. For definiteness we have not included the axial contribution
in $N_i$ and therefore in Eq.~(\ref{eq:eta2}). If included it would add
constructively in the denominator giving a smaller value of $\eta$.
From Eq.~(\ref{eq:albl}) we see
that the dependence of the loop contribution in $a_2 - 2 a_3$ is very 
mild and only $a_1$ is relevant. Values for $a_L$ and $b_L$ for fixed
$a_2 + 2 a_4$ are collected in Table 1. We notice that the dependence
on the combination $a_2 + 2 a_4$ is bigger in $a_L$.
\end{itemize}

\begin{table}
\begin{center}
\begin{tabular}{|c||c|c|c|c|c|c|}
\hline
& & & & & &  \\
$a_2 + 2 a_4$ & 0.2 & 0.4 & 0.6 & 0.8 & 1.0 & 1.2 \\
& & & & & &  \\
\hline
\hline
& & & & & &  \\
$a_L$ & $-$0.16 & $-$0.85 & $-$1.55 & $-$2.24 & $-$2.94 & $-$3.63 \\ 
& & & & & &  \\
\hline
& & & & & &  \\
$b_L$ & 0.85 & 0.98 & 1.12 & 1.26 & 1.40 & 1.54 \\
& & & & & &  \\
\hline
\end{tabular} 
\caption{Loop contributions $a_L$ and $b_L$ in 
Eqs.~(\ref{eq:m6oop},\ref{eq:albl}) for different values of the 
$a_2 + 2 a_4$ parameter as explained in the text.}
\end{center}
\end{table}

We have therefore three, a priori, free parameters~: $\rho$, $\omega$ and
$a_2 + 2 a_4$. Using now the experimental values for the 
branching ratio and the
slope of \kppg we can proceed to the analysis. However there is another
process where the breaking of the nonet symmetry in the weak sector could
play a r\^ole~: \kgg. A short note on this process is now needed.
\par
\kgg is a decay dominated by long-distance contributions which first
non vanishing amplitude is \ops in \chpt \cite{REFV,DH86}. Neglecting
CP violating effects the amplitude is defined by 
\begin{equation}
A\,[K_L(p) \rightarrow \gamma(q_1,\epsilon_1) \gamma(q_2,\epsilon_2)] \, 
= \, i \, A_{\gamma \gamma} \, \varepsilon_{\mu \nu \sigma \tau} \, 
\epsilon_1^{\mu}(q_1) \, \epsilon_2^{\nu}(q_2) \, q_1^{\sigma} \, 
q_2^{\tau} \; . 
\label{eq:klgg}
\end{equation}
There is, up to now, no full computation at \ops in \chpt. 
At present the amplitude is taken
\begin{equation}
A_{\gamma \gamma} \, = \, - \, \Frac{2}{\pi} \, G_8 \, F_{\pi} \, 
\alpha_{em}  \; F_2 \; , 
\label{eq:aggn}
\end{equation}
where
\begin{eqnarray}
F_2 \, & = & \, \Frac{1}{1-r_{\pi}^2} \, + \, 
\Frac{1}{3(1-r_{\eta}^2)} \, [ (1+\xi) \cos \theta + 2 \sqrt{2} \rho
\sin \theta ] \, \left[ \Frac{F_{\pi}}{F_8} \cos \theta - 
2 \sqrt{2} \Frac{F_{\pi}}{F_0} \sin \theta \right] \nonumber \\
& & \, - \, \Frac{1}{3(1-r_{\eta'}^2)} \, [ 2 \sqrt{2} \rho \cos \theta
- (1+\xi) \sin \theta ] \, \left[\Frac{F_{\pi}}{F_8} \sin \theta
+ 2 \sqrt{2} \Frac{F_{\pi}}{F_0} \cos \theta \right] \; ,
\label{eq:f2}
\end{eqnarray}
comes from the pole-like diagrams with the same definitions 
than $F_1$ in Eq.~(\ref{eq:m6f1}) \footnote{In Ref.~\cite{GZ97} has been
pointed out an extra contribution to $A_{\gamma \gamma}$ coming precisely
from a one loop evaluation using ${\cal L}_4^{|\Delta S|=1}$. We have not
been able to recover that result and found the claimed new contribution 
vanishes, therefore we do not take it into account. After reviewing
their calculation the authors of Ref.~\cite{GZ97} now agree with us.}. 
In our previous study on 
\kggs \cite{DP97a} we noticed that, keeping only the term in 
$F_2$ in Eq.~(\ref{eq:aggn}) with $\rho=1$ and $F_{\pi} = F_8 = F_0$
all the vector meson dominance models known were giving a wrong sign 
for the slope in 
\kggs. Due to the poor knowledge of the \kgg amplitude we think that the
problem could be the sign of this amplitude. Using the experimental
figure \cite{PDG96} $|A_{\gamma \gamma}^{exp}| = (3.51 \pm 0.05) \times
10^{-9} \, \mbox{GeV}^{-1}$, from Eq.~(\ref{eq:aggn}) we get
$|F_2^{exp}| \, = \, 0.883 \pm 0.013$. We have fixed the range 
$0 \leq \rho \leq 1$
by the conservative assumption that $|F_2^{theor}|$ does not differ
from the experimental figure more than a factor 2. We will include this 
process in our study and will combine, therefore, \kgg and \kppg.
\par
In our numerical analysis we use the following values for the known
parameters: the mixing angle between $\eta_8$ and $\eta_0$ that appears
in $F_1$ and $F_2$ is taken $\theta \simeq - 20^{\circ}$; the $SU(3)$
breaking parameter $\xi$ defined in Eq.~(\ref{eq:su3bre}) is 
$\xi \simeq 0.17$ \cite{DH86}; the ratios between the decay constants are 
$F_{\pi}/F_8 \simeq 0.77$ \cite{GL85} and $F_{\pi}/F_0 \simeq 0.98$ 
\cite{RO90} \footnote{These values are consistent with the recent
analysis of Ref.~\cite{VH97}.}; the couplings
defined in Eq.~(\ref{eq:vgpp}) are
$f_V \, = \, 2 g_V \, \simeq \, 0.18$, $h_V \, \simeq \, 0.037$ and
$\theta_V \, = \, 2 h_V$.
\par
Our analysis shows the following features~:
\begin{itemize}
\item[1/] There is a very mild dependence in the $\omega$ parameter for
the indicated domain $0.5 \leq \omega \leq 1.5$. It gives an uncertainty
in our predictions on \kppg of around $10\%$ for the branching ratio 
and $5 \%$ for the slope. Hereafter we take $\omega =1$ in our analysis.
\item[2/] In Table 2 we show, for fixed values of $a_2 + 2 a_4$, the 
region in $\rho$ that gives a reasonable result for 
$B(\kppge; E_{\gamma}^* > 20 \, \mbox{MeV})_{DE}$, the prediction for the
slope $c$ in \kppg and the values of $F_2$ in $A_{\gamma \gamma}$. We 
see that the branching ratio can always be accommodated for reasonable
values of $a_2 + 2 a_4$ and $\rho$. Taking into account $|F_2|$ in 
\kgg, however, there is a restriction on the lower and upper limits of
$a_2 + 2 a_4$, roughly $0.2 \leq a_2 + 2 a_4 \leq 1.2$ and also an 
internal region of this interval is excluded because $|F_2|$ is too small.
Thus the combined analysis of the experimental widths of \kppg and \kgg 
are consistent in two
regions of parameters that are ruled only by the value of the
combination $a_2 + 2 a_4$~:
\begin{eqnarray*} 
\mbox{\bf Region \, I} & : & \; \; 0.2 \, \leq \, a_2 + 2 a_4 \, 
\leq 0.6~, \; \; \; \; \;  \; 0 \, \leq \, \rho \, \leq \, 0.4~,  \\ 
& & \; \;  0.05 \, \leq \, \eta \, \leq \, 0.13~, \; \;  \; \; \; F_2 \, 
< \, 0~; \\
& & \\
\mbox{\bf Region \, II} & : & \; \; 0.8 \, \leq \, a_2 + 2 a_4 \, \leq 1.2~, \; 
\; \; \; \;  \; 0.6 \, \leq \, \rho \, \leq \, 1~,  \\
& & \; \; 0.18 \, \leq \, \eta \, \leq \, 0.23~, \; \;  \; \; \; F_2 \, > \, 0~.
\end{eqnarray*}
We remind that pion pole dominance in \kgg demands $F_2 > 0$. Our analysis, 
however,
shows that both possibilities are consistent in the framework we have 
developed here, and that the breaking of the nonet symmetry in the 
\opt weak vertex could consistently be as big as $\rho \simeq 0.2$.
Notice that if we had included the axial--vector contribution in the 
determination of 
$\eta$ in Eq.~(\ref{eq:eta2}) this parameter would diminish by a $15 \%$.
\end{itemize}

\begin{table}
\begin{center}
\begin{tabular}{|c|c|c|c|c|}
\hline
\multicolumn{1}{|c|}{$a_2 + 2 a_4$} &
\multicolumn{1}{|c|}{$\rho$} &
\multicolumn{1}{|c|}{$B(\kppge; E_{\gamma}^* > 20 \, \mbox{MeV})_{DE} 
\times 10^5$} &
\multicolumn{1}{|c|}{$c$} &
\multicolumn{1}{|c|}{$F_2$} \\
\hline
\hline
 & $-$0.05  & 4.09 & $-$1.52  & $-$1.89 \\
\cline{2-5} 
0.2 & 0.00 & 3.49 & $-$1.61 & $-$1.72 \\
\cline{2-5} 
& 0.05 & 2.93 & $-$1.71 & $-$1.55 \\
\hline
\hline
 & 0.15  & 4.00 & $-$1.56  & $-$1.22 \\
\cline{2-5} 
0.4 & 0.20 & 3.40 & $-$1.64 & $-$1.05 \\
\cline{2-5} 
& 0.25 & 2.85 & $-$1.74 & $-$0.89 \\
\hline
\hline
 & 0.35  & 3.93 & $-$1.59  & $-$0.55 \\
\cline{2-5} 
0.6 & 0.40 & 3.34 & $-$1.68 & $-$0.39 \\
\cline{2-5} 
& 0.45 & 2.80 & $-$1.78 & $-$0.22 \\
\hline
\hline
 & 0.55  & 3.85 & $-$1.62  & 0.11 \\
\cline{2-5} 
0.8 & 0.60 & 3.26 & $-$1.71 & 0.28 \\
\cline{2-5} 
& 0.65 & 2.72 & $-$1.82 & 0.45 \\
\hline
\hline
 & 0.75  & 3.78 & $-$1.65 & 0.79 \\
\cline{2-5} 
1.0 & 0.80 & 3.20 & $-$1.75 & 0.95 \\
\cline{2-5} 
& 0.85 & 2.67 & $-$1.85 & 1.11 \\
\hline
\hline
 & 0.95  & 3.70 & $-$1.68 & 1.45 \\
\cline{2-5} 
1.2 & 1.00 & 3.12 & $-$1.78 & 1.61 \\
\cline{2-5} 
& 1.05 & 2.60 & $-$1.89 & 1.78 \\
\hline
\end{tabular} 
\caption{Range of values of $\rho$ that give a reasonable prediction for
the branching ratio and the slope of \kppg at fixed value of $a_2 + 2 a_4$
for $\omega=1$.
The prediction for $F_2$ is also given. The experimental values are
$B(\kppge ; E_{\gamma}^* > 20 \, \mbox{MeV})_{DE} = (3.19 \pm 0.16) 
\times 10^{-5}$, $c_{exp} = -1.7 \pm 0.5$ and 
$|F_2| = 0.883 \pm 0.013$.}
\end{center}
\end{table}

\begin{itemize}
\item[3/] As can be seen in Table 2 the predicted slope $c$ of \kppg  is 
in a small range $c \simeq -1.6$, $-1.8$, independently from the range
of parameters and in excellent agreement with 
the experimental figure $c_{exp} = -1.7 \pm 0.5$. As seen from 
Eq.~(\ref{eq:cexpress}) the slope has two contributions~: vector exchange
and loops. The r\^ole of the latter is crucial in stabilizing the 
prediction. This can be seen in Table 3 where we show the values of the
slope including only the vector amplitude ($c \simeq -1.5$, $-2.8$) 
and the total result obtained
by adding the loop contribution ($c \simeq -1.6$, $-1.8$). 
The corrections to the width 
coming from the loop amplitudes are mainly relevant in Region II where
can reach even $100 \%$ for $a_2 + 2 a_4 = 1.2$. This is because in this
region there is an almost complete cancellation of the term 
$a_2 + 2 a_4 - F_1$ of $\widetilde{m}$ in Eq.~(\ref{eq:mtilde}) and
thus providing instability to the result; axial--vector exchange or 
higher ${\cal O}(p^8)$ corrections, not included in this analysis, might
then be relevant.
In Region I the loop contribution to the branching ratio is never bigger 
than $20 \%$.
\end{itemize}

\begin{table}
\begin{center}
\begin{tabular}{|c||c|c|c|c|c|c|}
\hline
& & & & & &   \\
$a_2 + 2 a_4$ & 0.2 & 0.4 & 0.6 & 0.8 & 1.0 & 1.2 \\
& & & & & &  \\
\hline
\hline
& & & & & & \\
$\rho$ & 0.0 & 0.2 & 0.4 & 0.6 & 0.8 & 1.0 \\
& & & & & &  \\
\hline
& & & & & & \\
$c$ (vectors only) & $-$1.54 & $-$1.70 & $-$1.89 & $-$2.13 & $-$2.43 & 
$-$2.82 \\ 
& & & & & &  \\
\hline
& & & & & &  \\
$c$ (vectors $+$ loops) & $-$1.61 & $-$1.64 & $-$1.68 & $-$1.71 & $-$1.75 &
$-$1.78 \\
& & & & & &  \\
\hline
\end{tabular} 
\caption{Comparison between the values of the slope $c$ of the magnetic
amplitude of \kppg for a fixed 
value of $a_2 + 2 a_4$ as given by vector exchange only and adding the loop
contributions in the total result. When the latter are included the prediction
is much more steady.}
\end{center}
\end{table}

In Fig.~7 we plot the spectrum of the {\em direct emission} 
\kppg in the photon energy ($E_{\gamma}^*$) for 
the two extreme values $a_2 + 2 a_4 = 0.2$ (Region I) with $\rho = 0$, 
$c = -1.61$, and $a_2 + 2 a_4 = 1.2$ (Region II) with $\rho=1$, 
$c = -1.78$. The tiny difference between both curves shows the model
dependence of our prediction.
We have normalized the spectrum to the 1937 events reported 
by the E731 experiment in Ref. \cite{RB93} that are also displayed. 
From this Fig.~7 it would seem that a bigger value of the magnitude of
the slope (and thus shifting the spectrum to lower values of $E_{\gamma}^*$)
might be needed. 
We would like to stress that the apparent disagreement between the data
and the theoretical predictions is due to the fact that data seem
to support a larger slope. However the form factor fitted by the
experimentalists gives a slope in very good agreement with our
prediction. A more accurate experimental determination is required (the 
events at low energy seem to have a huge error and thus a better
background subtraction from bremsstrahlung is needed).

\begin{figure}
\begin{center}
\leavevmode
\hbox{%
\epsfxsize=15cm
\epsffile{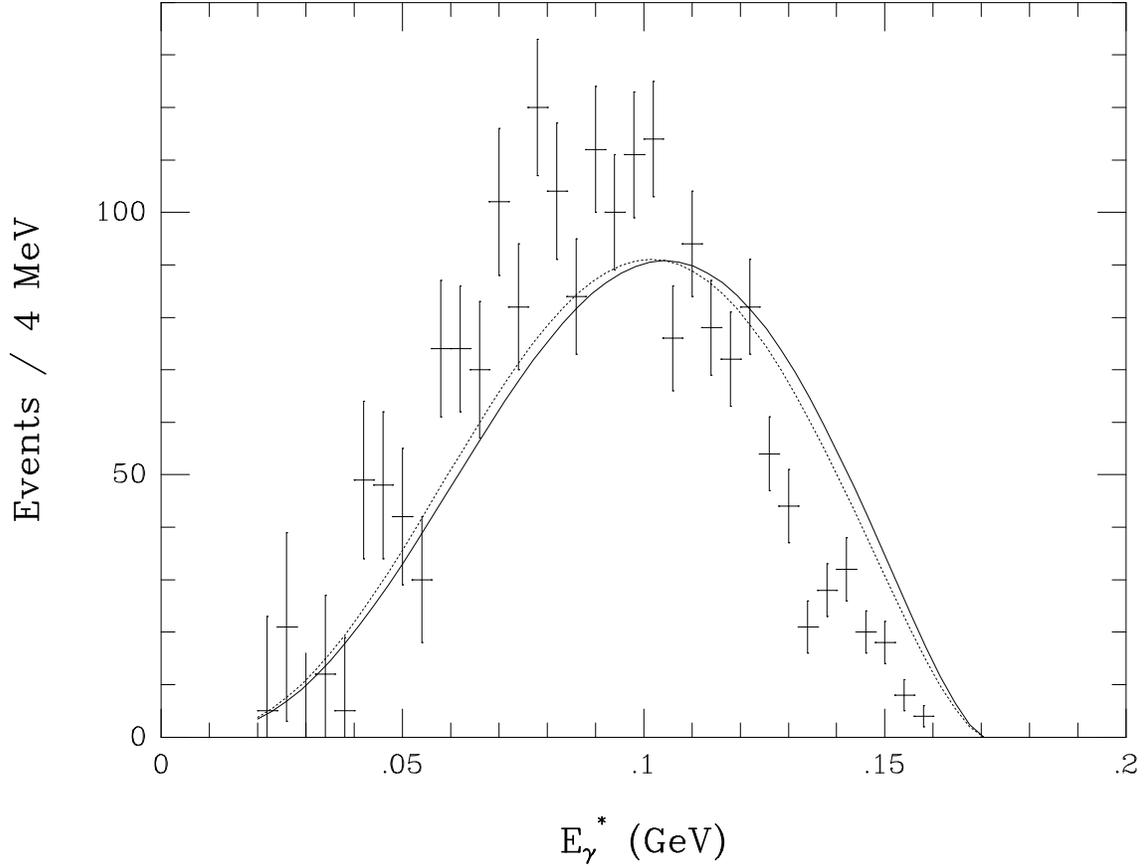}}
\end{center}
\vspace*{0cm}
\caption{Spectrum in $E_{\gamma}^*$ ($E_{\gamma}^* > 20 \, \mbox{MeV}$) of 
the $K_L \rightarrow \pi^+
\pi^- \gamma$ (DE) process normalized to the 1937 events of the E731
experiment of Ref.~\protect\cite{RB93} that are also displayed. 
The continuous line corresponds to $a_2 + 2 a_4 = 0.2$, $\rho=0$ and 
$\omega = 1$ with a predicted slope of $c = -1.61$. The dotted line
corresponds to $a_2 + 2 a_4 = 1.2$, $\rho=1$ and $\omega=1$ with 
a predicted slope of $c=-1.78$.}
\end{figure}

In Ref.~\cite{DH84} a parameter $f$ measuring the fraction of the 
$\Delta I = 1/2$ $K \rightarrow \pi \pi$ amplitude due to penguin 
effects was introduced. This ratio can be related with the parameter
$\rho$ introduced in Eq.~(\ref{eq:weakla}) that measures the breaking
of the nonet symmetry in the weak sector through the relation (assuming
CP conserved) $\rho = ( 3 f - 1 )/2$ \cite{DH84}. There are a lot of
uncertainties in the determination of $f$ and little can be said. We just
recall that for the generous range $0.1 \leq f \leq 0.8$ it implies
$-0.2 \leq \rho \leq 0.7$. Coming back to our analysis we see that
Region I implies $0.3 \leq f \leq 0.6$ while Region II allows
$0.7 \leq f \leq 1$ in the boundary of the reasonable range.
\par
A joint analysis of \kppg and \kgg has also been carried out previously
by other authors. Cheng \cite{CH90a,CH90b} only includes the term $a_2 + 
2 a_4 - F_1$ in $\widetilde{m}$ (Eq.~(\ref{eq:mtilde})) and, moreover, 
arbitrarily input $a_2 = a_4 = 1$. His conclusions are that the DE 
\kppg branching ratio excludes a small value for the $\rho$ parameter.
Ko and Truong \cite{KT91} and Picciotto \cite{PIC92} have considered
the contributions of the reducible anomalous amplitude (term in $F_1$)
and the {\em indirect} vector interchange (keeping, however, the whole
vector pole structure), but not the leading \opc amplitude~: they 
assume $a_2 = a_4 = 0$. These authors conclude that only a small breaking
of the nonet symmetry in the weak sector is compatible with their 
analysis ($\rho \simeq 0.8$, $0.9$) and therefore, that the $\pi^0$ pole
dominates the \kgg amplitude. Moreover their results give a small 
dependence of the magnetic amplitude in the photon energy and, consequently,
a too small slope $c$.
\par
All these previous analyses emphasize that only our Region II is allowed. 
In our more complete analysis we show that this is not the case and both
solutions are kept open. In fact several features (the sign of the slope
in \kggs and the ratio $f$) lead to the scenario described by the 
parameters in Region I.
\par
A comment about the convergence of the chiral expansion is also interesting.
The bulk of the contribution to the width of \kppg comes from local
contributions at \opc and \ops, but the loop amplitudes are relevant
for the region $0.8 \leq a_2 + 2 a_4 \leq 1.2$. The slope in 
$E_{\gamma}^*$ starts at 
\ops in \chpt and it is mostly due to the vector exchange but the loop
contribution can be as big as $30 \%$ and in any case is crucial to 
give a steady prediction. Local \ops corrections due to resonance exchange
(other than vectors), even if suppressed by heavier masses, could give a 
non--negligible contribution to the width if numerical factors overcome 
the suppression. The analysis of ${\cal O}(p^8)$ is 
very much cumbersome. It is curious to notice, however, that the 
${\cal O}(p^8)$ (and higher orders) amplitude specified in Eq.~(\ref{eq:m68})
(that we have not included in our analysis) does not seem smaller than the 
pure \ops $M_{ind}^{(6)}$ in Eq.~(\ref{eq:m6ind}) because, in principle,
$F_1 \sim {\cal O}(1)$. 
If the chiral expansion has to converge one would
expect either a cancellation with other ${\cal O}(p^8)$ contributions
or that the parameters allow a self--suppression. Hence 
for our preferred values of $a_2 + 2 a_4$, 
$\rho = 0.8$ implies $F_1 \simeq 0.96$
but $\rho = 0.2$ gives $F_1 \simeq -0.07$ that is one order of magnitude
smaller. This fact together with the
statement pointed out above that in Region II the \ops loop corrections
to the width are huge indicates
that a better convergence of the chiral expansion (in the terms considered
here) seems to be achieved with $a_2 + 2 a_4$ and $\rho$ small
\footnote{We have
studied the scale dependence of the loop amplitude. As a matter of fact
this is not negligible by itself (i.e. $\sim 30 \%$ in the loop for
a reasonable range of variation of $\mu$). However
the loop contributions play a very marginal role in the observables
(i.e. rate and spectrum) for Region I; thus this change is negligible.
This is not the case in Region II (where strong cancellations make the
final result sensitive to any small change) confirming our conclusion
that predictions in this region are rather loose.}.

\section{Conclusion}
\hspace*{0.5cm}
The measured energy dependence of the spectrum for the process 
\kppg shows that the slope of the magnetic amplitude defined in 
Eq.~(\ref{eq:slope}) has a relative big size thought to be due to
a relevant vector meson exchange contribution at \ops in \chpt.
Moreover with our present knowledge of the phenomenology the width of 
the process cannot be predicted in a model independent way.
In Ref.~\cite{DP97a} we pointed out that the sign of the
\kgg amplitude has to be opposite to the one predicted by pion pole
dominance in this decay. Both processes are correlated because receive
a reducible contribution depending on the unknown $\rho$ parameter
(Eq.~(\ref{eq:weakla})) that measures the breaking of the nonet symmetry
in the \opt weak vertex.
\par
In this work we have presented a systematic study of the magnetic amplitude
of the process \kppg up to \ops in \chpt. We have computed the local
contributions (due to a reducible anomalous amplitude and vector meson
exchange) and the one--loop magnetic amplitude at this order.
\par
The \opc magnetic amplitude ($M^{(4)}$) and part of the loop contribution
($M_{N_i}^{(6)}$) are model dependent because the couplings $N_i$ in 
${\cal L}_4^{|\Delta S|=1}$ Eq.~(\ref{eq:weak4}) are not known from the
phenomenology. We have included the spin--1 resonance exchange contribution 
to the relevant $N_i$ couplings that we have evaluated in the approach 
described in Ref.~\cite{DP97b}. 
Moreover part of the vector exchange amplitudes due to
diagrams in Fig.~4 where a weak $VP\gamma$ or $VPP$ appear are also
not known. Driven by the good description we have achieved in 
Ref.~\cite{DP97a} for the processes $K \rightarrow \pi \gamma \gamma$ and
$K_L \rightarrow \gamma \ell^+ \ell^-$ we have used the Factorization 
Model in the Vector Couplings (FMV) \cite{DP97a,DP97b} in order to provide
the {\em direct} weak $VP\gamma$ and $VPP$ vertices.
\par
The conclusions of our analysis can be stated as follows~:
\begin{itemize}
\item[1/] We are able to accommodate the experimentally determined branching 
ratio
$Br(\kppge ; E_{\gamma}^* > 20 \, \mbox{MeV})_{DE}$ inside a reasonable
range of variation of the parameters $a_2 + 2 a_4$ and $\rho$. When 
combined with the \kgg amplitude we find two regions for the 
combination $a_2 + 2 a_4$~: $0.2 \leq a_2 + 2 a_4 \leq 0.6$ and 
$0.8 < a_2 + 2 a_4 \leq 1.2$. Away of this range $|A_{\gamma \gamma}|$
is predicted to differ for more than a factor two of the experimental
figure.
Remarkably in any of these cases the value given by Eq.~(\ref{eq:eta2})
is $\eta \leq 0.3$ implying, therefore, that the octet operators are 
not enhanced and are consistent with the perturbative expectation provided by
the Wilson coefficient ($\eta \simeq 0.2$). This we can interpret as a 
fiducial behaviour of factorization showing that the dynamical features
introduced in our analysis through this hypothesis are compatible with the
perturbative expansion that becomes predictive a this point.
\par
We have also concluded that the Region II is perhaps sensitive to effects
not included here like \ops axial--vector exchange or non--resonant 
contributions and ${\cal O}(p^8)$ corrections.
\item[2/] Independently of the parameter variation we consistently predict
a small range for the slope of \kppg, roughly $c \, \simeq \, -1.6$, 
$-1.8$ in very good agreement with the experimental result $c_{exp} 
= -1.7 \pm 0.5$. The loop contributions are shown to be crucial to 
stabilize the slope in this small range.
As a consequence the spectrum plotted in Fig.~7 is a pure prediction
of our framework. Corrections to it can come from other resonance interchange
at \ops (axial-vectors, scalars, ...) or ${\cal O}(p^8)$ effects.
\item[3/] From Table 2 we see that, in our joint analysis, both 
signs of the \kgg amplitude are consistent with the phenomenology, in
contradistinction to the conclusion reached by previous studies. Hence
the result $F_2 <0$ (pion pole dominance implies $F_2 >0$) is allowed
in consistency with our statement in Ref.~\cite{DP97a} that the slope
in \kggs, experimentally determined, would imply that sign. If this is
the case we would conclude that, in opposition to previous results
\cite{CH90b,KT91,PIC92}, a bigger breaking of the nonet symmetry in the weak
vertex is called for ($\rho \simeq 0.2$). Of course, 
at this point, it cannot be excluded a solution with $\rho \simeq 0.8$, 
and blame the change of sign of the \kgg 
amplitude to a, still unknown, huge higher order correction.
\end{itemize}

\par
The relevance of other resonance exchange generated \ops or higher order 
contributions would be very much clarified once a more accurate 
measurement of the spectrum in $E_{\gamma}^*$ is carried out. The experiment
KLOE at DA$\Phi$NE expects to have $35,000$ \kppg events p/year due to DE 
with $E_{\gamma}^* > 20 \, \mbox{MeV}$ \cite{DE95} and therefore should be
able to improve statistics and accuracy in this channel. Other two new 
experiments NA48 at CERN and KTEV at Fermilab could also bring new features
on this process. These are going to focus on the related 
$K_L \rightarrow \pi^+ \pi^- \gamma^*$ process where to keep under theoretical
control the leading real photon channel is crucial to disentangle 
its $q^2$ dependence.
\vspace*{1cm} \\
{\bf Acknowledgements}
\vspace*{0.5cm}\\
\hspace*{0.5cm}
The authors wish to thank F.J. Botella, F. Cornet, G. Ecker, G. Isidori,
H. Neufeld and A. Pich for interesting and fruitful discussions on the 
topics of this paper. J.P. is partially supported by Grant PB94--0080 of 
DGICYT (Spain) and Grant AEN--96/1718 of CICYT (Spain).

\newpage

\appendix
\newcounter{pla}
\renewcommand{\thesection}{\Alph{pla}}
\renewcommand{\theequation}{\Alph{pla}.\arabic{equation}}
\setcounter{pla}{1}
\setcounter{equation}{0}

\section*{Appendix A~: Kinematics on \kppg}
\hspace*{0.5cm}
In Section 3 we defined the amplitudes and wrote explicitly the
differential cross section for an unpolarized photon in \kppg. 
Here, for completeness, we collect some useful kinematical relations 
\cite{DI95,DM92}.
\par
In \kppg the most useful variables are~: i) the photon energy in the
kaon rest frame ($E_{\gamma}^*$), ii) the angle ($\theta$) between the 
$\gamma$ and $\pi^+$ momenta in the di--pion rest frame. The relations
between $(E_{\gamma}^*, \theta)$ and the $z_i$ defined in 
Eq.~(\ref{eq:defz}) are
\begin{equation}
z_{\pm} \, = \, \Frac{E_{\gamma}^*}{2 m_K} \, ( \, 1 \, 
\mp \, \beta \cos \theta \, ) \;  , \; \; \; \; \; \; \; \; \; \; \; \; \; 
\; z_3 \, = \, \Frac{E_{\gamma}^*}{m_K} \; , 
\label{eq:a1}
\end{equation}
where $\beta = \sqrt{1-4 m_{\pi}^2/(m_K^2 - 2 m_K E_{\gamma}^*)}$. The
kinematical limits on $E_{\gamma}^*$ and $\theta$ are given by
\begin{equation}
0 \, \leq \, E_{\gamma}^* \, \leq \, \Frac{m_K^2 - 4 m_{\pi}^2}{2 m_K}
\; , \; \; \; \; \; \; \; \; \; \; \; \; \; \; \; \; \; -1 \, 
\leq \, \cos \theta \, \leq \, 1 \; . 
\label{eq:a2}
\end{equation}
The differential rate in terms of these variables for an unpolarized 
photon is
\begin{equation}
\Frac{\partial^2 \Gamma}{\partial E_{\gamma}^* \, \partial \cos \theta}
\, = \, \Frac{(E_{\gamma}^*)^3 \beta^3}{512 \pi^3 m_K^3} \, 
\left( 1 - \Frac{2 E_{\gamma}^*}{m_K} \right) \sin^2 \theta \, 
\left( \, |E|^2 \, + \, |M|^2 \, \right) \;  .
\label{eq:a3}
\end{equation}

\appendix
\newcounter{quijote}
\renewcommand{\thesection}{\Alph{quijote}}
\renewcommand{\theequation}{\Alph{quijote}.\arabic{equation}}
\setcounter{quijote}{2}
\setcounter{equation}{0}

\section*{Appendix B~: The weak VPP vertex in the Factorization Model
in the Vector couplings}
\hspace*{0.5cm}
Following the same procedure that we used in Ref.~\cite{DP97a} to construct
the weak $VP\gamma$ vertex we proceed to evaluate the weak VPP vertex
for \kppg.
\par
The bosonization of the $Q_{-}$ operator in Eq.~(\ref{eq:qm}) can be 
carried out in the FMV from the strong effective action $S$ of a chiral
gauge theory. For later use let us split the strong action and the
left-handed current into two pieces~: $S = S_1 + S_2$ and 
${\cal J}_{\mu} \, = \, {\cal J}_{\mu}^1 + {\cal J}_{\mu}^2$, respectively.
Then in the factorization approach the $Q_{-}$ operator is represented
by
\begin{equation}
Q_{-} \, \leftrightarrow \, 4 \, \left[ \, \langle \, \lambda
\, \{ \, {\cal J}_{\mu}^1, {\cal J}^{\mu}_2 \, \} \, \rangle \, - \, 
\langle \, \lambda \, {\cal J}_{\mu}^1\, \rangle \, \langle \, 
{\cal J}^{\mu}_2 \, \rangle \, - \, 
\langle \, \lambda \, {\cal J}_{\mu}^2\, \rangle \, \langle \, 
{\cal J}^{\mu}_1 \, \rangle \, \right] \;  , 
\label{eq:c1}
\end{equation}
with $\lambda \equiv (\lambda_6 - i \lambda_7)/2$ and, for generality,
the currents have been supposed to have non--zero trace.
\par
In order to apply this procedure to construct the factorizable 
contribution to the weak ${\cal O}(p^3)$ VPP vertex we have to identify
in the full strong action the pieces (at this chiral order) that can
contribute. It turns out that there are four terms in the strong 
action that can play a r\^ole. Analogously to the specified procedure
we define correspondingly~:
\begin{eqnarray}
S \, & = & S_V \, + \, S_P \; , \nonumber \\
S_V \, & = & S_{V\gamma} \, + \, S_{VPP} \;  , 
\label{eq:c2} \\
S_P \, & = & S_2^{\chi} \, + \, S_4^{\chi} \;  , \nonumber 
\end{eqnarray}
where the notation is self-explicative and the actions correspond to 
the lagrangian densities proportional to $f_V$ and $g_V$ in ${\cal L}_{V}$ 
Eq.~(\ref{eq:vgpp}), 
${\cal L}_2$ in Eq.~(\ref{eq:str2}) and ${\cal L}_4$ in Eq.~(\ref{eq:op4l9}). 
\par
Evaluating the left-handed currents and keeping only the terms of 
interest we get
\begin{eqnarray}
\Frac{\delta S_V}{\delta \ell^{\mu}} \, &  = &  \, 
\Frac{f_V}{\sqrt{2}} \, \left[ \partial^{\alpha} u^{\dagger} \, 
V_{\alpha \mu} \, u \, + \, u^{\dagger} \, V_{\alpha \mu} \, 
\partial^{\alpha} u \, \right] \, + \, i \, 
\Frac{g_V}{\sqrt{2}} \, u^{\dagger} \, [ V_{\alpha \mu} , u^{\alpha} ] \, 
u \; , \nonumber \\
& & \label{eq:c3} \\
\Frac{\delta S_P}{\delta \ell^{\mu}} \, & =  & \, 
- \, \Frac{F_{\pi}^2}{2} \, u^{\dagger} \, u_{\mu} \, u \, + \, 
i \, L_9 \, \partial^{\alpha} \left( \, u^{\dagger} \, 
[u_{\alpha},u_{\mu}] \, u \right) \;  .
\nonumber
\end{eqnarray}
Then the effective action in the factorizable approach is
\begin{eqnarray}
{\cal L}_W^{fact}(VPP) \, & = & \, 4 \, G_8 \, \eta_{VPP} \, 
\left[ \, \langle \, \lambda \, \left\{ \Frac{\delta S_V}{\delta \ell^{\mu}} \,
, \, \Frac{\delta S_{P}}{\delta \ell_{\mu}} \, \right\} \rangle
\, - \, \langle \lambda \Frac{\delta S_V}{\delta \ell^{\mu}} \rangle
\langle \Frac{\delta S_{P}}{\delta \ell_{\mu}} \rangle \right.
\nonumber \\
& & \; \; \; \; \; \; \; \; \; \; \; \; \; \; \; \; \; \; \left.
\, - \, \langle \lambda \Frac{\delta S_{P}}{\delta \ell^{\mu}} 
\rangle \langle \Frac{\delta S_V}{\delta \ell_{\mu}} \rangle \, \right] 
\; + \, h.c. \; \; . 
\label{eq:c4} 
\end{eqnarray}
There is however an extra contribution that remains to be taken into 
account. In Ref.~\cite{DP97b} we have shown that in the conventional vector 
formulation
one can construct a weak ${\cal O}(p)$ coupling involving vectors as 
\begin{equation}
{\cal L}_{{\cal O}(p)}^V \, = \, 
G_8 \, F_{\pi}^4 \, \left[ \, \omega_1^V \, 
\langle \, \Delta \, \{ V_{\mu}, u^{\mu} \} \, \rangle \, + \, 
\omega_2^V \, \langle \, \Delta u_{\mu} \, \rangle \, \langle V^{\mu} 
\rangle \, \right]  \; , 
\label{eq:c5}
\end{equation}
where the couplings $\omega_i^V$ are in the FM
\begin{equation}
\omega_1^V \, = \, \sqrt{2} \, \Frac{m_V^2}{F_{\pi}^2} \, 
f_V \, \eta_V \; , \; \; \; \; \; \; \; \; \; \; \; \; \; \; \; 
\omega_2^V \, = \, -  \, \omega_1^V \; , 
\label{eq:c6}
\end{equation}
with $\eta_V$ the unknown factorization factor.
If one rotates away these terms through the shift
\begin{equation}
V_{\mu} \, \rightarrow \, V_{\mu} \, - \, 
\Frac{G_8 F_{\pi}^4}{m_V^2} \, \left[ \, 
\omega_1^V \, \{ \Delta, u_{\mu} \} \, + \, \omega_2^V \, 
\langle \Delta u_{\mu} \rangle \, \, \right] \; ,
\label{eq:c7}
\end{equation}
the kinetic term of the vector mesons 
\begin{equation}
{\cal L}_K \, = \, - \, \Frac{1}{4} \, \langle \, V_{\mu \nu} \, 
V^{\mu \nu} \, \rangle \, + \, \Frac{m_V^2}{2} \, \langle \, 
V_{\mu} \, V^{\mu} \, \rangle~,
\label{eq:c8}
\end{equation}
generates also an ${\cal O}(p^3)$ weak VPP coupling. 
In Eq.~(\ref{eq:c8}) $V_{\mu \nu}$ has been defined in connection 
with Eq.~(\ref{eq:vgpp}).
\par
Comparing the full lagrangian density ${\cal L}_{W}(VPP)$ in 
Eqs.~(\ref{eq:mostgvpp},\ref{eq:fullcurvpp},\ref{eq:svpp}) we find
the predictions of the FMV model for the couplings $\sigma_i$. 
Using $f_V \, = \, 2g_V$ we get
\begin{eqnarray}
\sigma_1^{FMV} \, & = & \; \; \; \; \, \Frac{f_V}{\sqrt{2}} \, 
\left[ \, \eta_V \, + \, 4 \, L_9 \, \Frac{m_V^2}{F_{\pi}^2} \, \eta_{VPP}
\, \right]  \; , \nonumber \\
& & \nonumber \\
\sigma_2^{FMV} \, & = & \, -  \, \sqrt{2} \, f_V \, \eta_{V} 
\; , \label{eq:c9} \\
& & \nonumber \\
\sigma_3^{FMV} \, & = & \, - \,  \sqrt{2} \, f_V \, 
\left[ \, \eta_V \, + \, 2 \, L_9 \, \Frac{m_V^2}{F_{\pi}^2} \,  
\eta_{VPP} \, \right] \; . \nonumber 
\end{eqnarray}

\appendix
\newcounter{hamlet}
\renewcommand{\thesection}{\Alph{hamlet}}
\renewcommand{\theequation}{\Alph{hamlet}.\arabic{equation}}
\setcounter{hamlet}{3}
\setcounter{equation}{0}

\section*{Appendix C~: Loop integrals}
\hspace*{0.5cm}
Here we collect the functions appearing in the loop contributions
to the magnetic amplitude.
\par
The function $K[q^2,m_i^2,m_j^2]$ appearing in the loop contributions
to the magnetic amplitude $M^{(6)}_{WZW}$ Eq.~(\ref{eq:m6wzw}) and 
$M^{(6)}_{N_i}$ Eq.~(\ref{eq:m6ni}) is defined through
\begin{equation}
\int \, \Frac{d^D \ell}{(2\pi)^D} \, 
\Frac{\ell^{\mu} \ell^{\nu}}{[\ell^2 - m_i^2][(\ell-q)^2-m_j^2]} \, 
= \, i \, g^{\mu \nu} \, q^2 \, B_{22}[q^2, m_i^2,m_j^2] \, + \, ...\; , 
\label{eq:b3}
\end{equation}
as
\begin{equation}
q^2 \, B_{22}[q^2,m_i^2,m_j^2] \, |_{\overline{MS}} \, = \, 
\left( \Frac{m_K}{4 \pi} \right)^2 \, K[q^2, m_i^2,m_j^2] \; , 
\label{eq:b4}
\end{equation}
where in the $\overline{MS}$ scheme the term proportional to 
 $\lambda_{\infty} \, = \, \Frac{2}{D-4} - ( \Gamma'(1) + 
\ln (4\pi) + 1 )$ appearing in $B_{22}[q^2,m_i^2,m_j^2]$ has been 
subtracted.

The $F[q^2,m^2]$ function defined in the same loop amplitudes is
\begin{equation}
F\,[q^2,m^2] \, = \, 
\left( \, 1 \, - \, \Frac{x}{4} \, \right) \, 
\sqrt{1 \, - \, \Frac{4}{x}} \, 
\ln \left( \Frac{\sqrt{x-4} \, + \, \sqrt{x}}{\sqrt{x-4} \, - \, 
\sqrt{x}} \right) \, - \, 2 \;  ,
\label{eq:b5}
\end{equation}
with $x=q^2/m^2$ (note that this function is usually quoted in the
literature \cite{BB90} with an extra factor of $m^2$).
\par
For equal masses the $K$ and $F$ functions are related as
\begin{equation}
K[q^2,m^2,m^2] \, = \, - \, \Frac{1}{12 m_K^2} \, \left[ \, 
\left( \, 6 \, m^2 \, - \, q^2 \, \right) \, 
\ln \left(\Frac{m^2}{\mu^2}\right) \, + \, \Frac{5}{3} \, q^2 \, + \, 
4 \, m^2 \, F[q^2,m^2] \, \right]~.
\label{eq:b6}
\end{equation}

\end{document}